\documentclass[letterpaper, 10 pt, conference]{ieeeconf}  % Comment this line out if you need a4paper

\IEEEoverridecommandlockouts                              % This command is only needed if 

\usepackage{graphicx}      % include this line if your document contains figures
\usepackage{caption}
\usepackage{subcaption}
\usepackage{amsmath}
\usepackage{amssymb}
\usepackage{xcolor}
\usepackage{import}
\newtheorem{theorem}{Theorem}

\newtheorem{remark}{Remark}
\newtheorem{proposition}{Proposition}
\usepackage{multirow}

\usepackage{algorithm2e}
\usepackage{mathtools, cases}
\usepackage{cuted} 
\newcommand{\sign}{\text{sign}}
\newcommand{\sat}{\text{sat}}
\usepackage{placeins}
\usepackage{bm}

\makeatletter
\let\NAT@parse\undefined
\makeatother
\usepackage[hidelinks]{hyperref} % <====================== last packge to be called

\title{\LARGE \bf
Switched Vector Field-based Guidance for General Reference Path Following in Planar Environment} %Using the Switched Vector Field Method for Faster Convergence with Realistic Turn Rate}

\author{Subham Basak$^{1}$ and Satadal Ghosh$^{2}$% <-this % stops a space
\thanks{The authors are with the Department of Aerospace Engineering, Indian Institute of Technology Madras, Chennai-600036, India,
        {\tt\small ae20d412@smail.iitm.ac.in, satadal@iitm.ac.in}}%
}

\begin{document}

\maketitle
\thispagestyle{empty}
\pagestyle{empty}

%%%%%%%%%%%%%%%%%%%%%%%%%%%%%%%%%%%%%%%%%%%%%%%%%%%%%%%%%%%%%%%%%%%%%%%%%%%%%%%%
\begin{abstract}

Reference path following is a key component in the functioning of almost all engineered autonomous agents. Among several path following guidance methods in existing literature, vector-field-based guidance approach has got wide attention because of its simplicity and guarantee of stability under a broad class of scenarios. However, the usage of same cross-track-error-dependent structure of desired vector field in most of the existing literature irrespective of instantaneous cross-track error and course angle of unmanned vehicle makes it quite restrictive in attaining faster convergence and also leads to infeasibly high turn rate command for many scenarios. To this end, this paper presents a novel switched vector field-based guidance for following a general reference path, in which the structure of the desired vector field depends on instantaneous cross-track-error and vehicle's course angle. While the developed method ensures faster convergence, it also ensures that the guidance command always stays within a realistic threshold satisfying its curvature constraint, thus making it more real-life implementable for autonomous vehicles with kino-dynamic constraints. Theoretical analysis for convergence of the developed guidance scheme is presented. Possibilities of undesirable chattering at phase transitions are also eliminated. Numerical simulation studies are presented to validate the satisfactory performance of the developed algorithm. 
% Finally, an illustration is also provided to demonstrate the performance of a preliminary extension of the developed guidance scheme in following a reference circular path.

\end{abstract}

%%%%%%%%%%%%%%%%%%%%%%%%%%%%%%%%%%%%%%%%%%%%%%%%%%%%%%%%%%%%%%%%%%%%%%%%%%%%%%%%
\section{INTRODUCTION}

%Over last two decades, development of unmanned aerial vehicles (UAVs) has gained significant momentum due to the rapid advances in technologies related to sensors, vehicle frames, electronics and controls. As a direct consequence, UAVs have seen large-scale adoption in different civilian and military applications. %In all these applications UAVs are assigned to perform certain desirable tasks. 
%In most of these scenarios, UAVs are required to autonomously follow pre-defined reference paths as precisely as possible.

With the steady advances in technologies related to autonomy, unmanned vehicles (UxVs) have seen large-scale adoption in different civilian and military applications. Capability of autonomously following pre-defined reference paths is a key component in most of these scenarios.

In existing literature, path following problem has been addressed in three broad ways - control theoretic, guidance theoretic and vector field-based methods \cite{sujit}. Among control-theoretic methods, linear controllers like proportional–integral–derivative (PID) \cite{rhee} and optimal control-based methods \cite{Sujit_LQR}, \cite{Tsourdos_Optimal_jgcd_2019} on linearized geometries have been presented for developing path following control schemes. %Optimal control-based formulations have also been presented in linearlized framework of nonlinear path following problem \cite{Tsourdos_Optimal_jgcd_2019}. 
However, these methods majorly rely on small deviation from nominal assumption that is valid for sufficiently small heading error or cross-track error geometries. Lyapunov-based nonlinear controllers have also been developed in \cite{Nonlin_Control_Lyapunov_jgcd_2013}, \cite{Lozano_Nonlin_icuas_2013} for regulation of path following error. While these nonlinear control techniques possess guaranteed stability and reliable tracking, the guidance commands are complicated, model-dependent and also depend on the magnitude of error \cite{Park_DiffGeometry_jgcd_2015}, which was resolved in 
differential geometry-based path following guidance formulations in \cite{Park_DiffGeometry_jgcd_2015}, \cite{Gates_DiffGeometry_jgcd_2010}. However, their convergence proofs were restricted to constant-curvature planar paths.
% Similarly, nonlinear differential geometry-based path following guidance formulations were presented in \cite{Park_DiffGeometry_jgcd_2015}, \cite{Gates_DiffGeometry_jgcd_2010}.
On the other hand, guidance-theoretic path following methods are model-independent, simple and easy to implement. Some of these schemes involve Pure pursuit (PP) \cite{PP_jgcd_2010}, \cite{PP_imecheg_2014}, line-of-sight (LOSG) guidance \cite{Rysdyk_LOSG_jgcd_2006}, 
trajectory shaping guidance \cite{Ratnoo_TrajShape_jgcd_2015},
combination of PP and LOSG \cite{kothari2014uav}. Apart from these, A Proportional Navigation (PN)-resembling guidance was devised in \cite{park2007} for following a virtual target at a fixed-look-ahead distance on a reference path (straight line or circular or perturbed path), while another PN-based guidance was formulated in \cite{Dhananjay_PN_Look-ahead} for following a virtual target at a fixed look-ahead direction. % A guidance strategy resembling Proportional Navigation (PN) philosophy was devised in \cite{park2007} for following a virtual point at a fixed look-ahead distance ($L_1$) on the desired straight line or circular or perturbed path. Another PN-based guidance strategy was developed in \cite{Dhananjay_PN_Look-ahead} to follow a virtual target at a fixed look-ahead distance at a desired direction depending on the geometry of the reference path. 
Apart from these, Dubins curve-based guidance scheme for following an optimal feasible path in three-dimensional space was presented in \cite{Sikha_Dubins_3D}. Among all these nonlinear formulations, the look-ahead-based path following guidance \cite{park2007} has got popularity because of its simplicity and ease in application, asymptotic stability and robustness. However, its drawback of constant look-ahead distance makes it restrictive for realistic path following applications. To this end, a radius of reference path curvature-dependent variable look-ahead distance guidance was presented in \cite{SG_Variable_L1_icuas}. The major limitation of the look-ahead-based guidance schemes is that the initial position of the UxV needs to be inside the range of specified look-ahead distance from the reference path, thus necessitating an additional mid-course guidance law if the initial position is outside of the look-ahead distance.

% \textcolor{blue}{Among nonlinear methods, look-ahead path guidance \cite{park2007} gained popularity for its simplicity, stability, but had a fixed look-ahead distance drawback. A variable distance based on path curvature was proposed in \cite{SG_Variable_L1_icuas}.The main limitation of look-ahead-based guidance is that the UAV must start within the specified look-ahead distance from the reference path, requiring a mid-course guidance law if it starts beyond this distance.}
In this context, Vector field-based methods \cite{nelson2007}, are gainfully leveraged due to their low cross-track error and low control effort compared to other existing path following algorithms \cite{sujit}. Following this approach, the desired course angle of the UxV is constructed using a vector field, whose integral curves asymptotically direct the UxV to the pre-determined reference path. Stability analysis of the vector field-based path-following method for reference straight and circular paths was presented in \cite{nelson2007}, while this method was extended in \cite{griffiths2006vector} to following a general reference path with small straight line approximation. 
% using an \textcolor{red}{adaptive} vector field-based approach.
In \cite{lawrence2008}, a notion of circular attractor Lyapunov function was defined to generate the vector field for a reference path following. Distance-based application of irrotational and rotational parts of a vector field was utilized in \cite{liang2016combined} for path following in 3-D. A general method for creating the vector field needed for navigation in n-dimensional space was presented in \cite{goncalves2010vector}, singularity problem of which was subsequently addressed in \cite{yao2021singularity}.

However, in most of the vector field-based methods in existing literature, the field is defined as a function of %the distance of the UxV from the reference path, termed as 
only cross-track error, thus making the vector field form same irrespective of the magnitude of cross-track error. This significantly limits the possibility of enhanced convergence to the reference path. Moreover, in these formulations, as the cross-track error approaches to infinity, the commanded desired heading angle approaches to a constant, which often leads to kino-dynamically infeasible high turn rate command if the course angle error is very high. %lies in one half-space of the angular spectrum. 
%These, in effect, restrict the application of such vector field formulations in real vehicles. 
Additionally, curvature constraints play a pivotal role in achieving precise trajectory tracking and optimizing vehicle performance while taking real-world limitations into account \cite{ratnoo2023arcsineVF}. %In \cite{ratnoo2023arcsineVF}, a vector field approach for both straight-line and circular paths is introduced, providing a mathematical expression for the maximum curvature. However, it's important to note that the maximum curvature condition applies exclusively to the straight-line path following scenario.}
To address all these considerations, in this paper, a general reference path following guidance scheme for planar applications is presented based on a novel formulation of switched vector field, in which its dependence on cross-track error facilitates in augmented convergence of the developed guidance scheme, while its dependence on UAV course ensures that the turn rate command remains within a kino-dynamically feasible threshold. Stability analysis for finite time convergence by the developed guidance strategy is given. Thus, attaining a faster path following while maintaining the guidance command within feasible threshold value forms the salient feature of the developed guidance scheme.

The rest of this paper is organized as follows: Section \ref{sec:problem_defination} defines the problem. Section \ref{sec:line_VF} introduces different vector fields and develops as well as analyzes a novel guidance strategy for a following general reference path. %The proposed guidance laws are analyzed, and necessary theoretical guarantees are provided in this section. 
Section \ref{sec:Curvature_Constraints} delves into the curvature limitations associated with general reference path following. The proposed guidance law is then validated using numerical simulations in Section \ref{sec:Simulation_results}. Finally, Section \ref{sec:conclusions} concludes the overall guidance algorithm. 

% \textcolor{green}{Cover more area across the path accurately}

\section{Problem Definition} \label{sec:problem_defination}
\subsection{Navigation Dynamics}
\begin{figure}[]
    \centering
    \includegraphics[scale=.55]{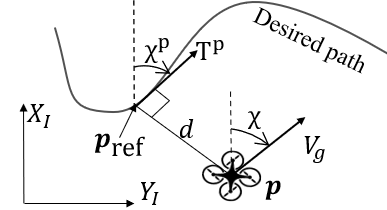}
    \caption{General reference path following geometry}
    \label{fig:VF_for_line_following}
\end{figure}

Consider the problem of following a reference path as shown in Fig.~\ref{fig:VF_for_line_following}. The navigational dynamics of a constant speed UAV flying at a constant altitude can be stated as,
\begin{align}%\label{eq:dynamics}
    \Dot{\chi} = V_g \cos(\chi);\quad \Dot{y} = V_g \sin(\chi);\quad \Dot{\chi} = \alpha(\chi^c-\chi) \label{eq:dynamics_chi_dot}
\end{align} 
Here, $\bm{p}=[x,y]^T$ denotes the UAV's position w.r.t. inertial reference frame $(O-X_I,Y_I)$. Its ground speed and course angle are given as $V_g = \sqrt{(V_{a_x}+W_x)^2 + (V_{a_y}+W_y)^2}$ and $\chi = \tan^{-1}((V_{a_y}+W_y)/(V_{a_x}+W_x))$, respectively, where $V_a$, $W$ denote the UAV airspeed and wind speed, respectively. Subscripts $x$ and $y$ denote along $X_I$ and $Y_I$ directions.
% expressed as,
% \begin{align}
%     & V_g = \sqrt{V_a^2+V_w^2}
%     \\& \chi = \textcolor{red}{???????}
% \end{align}
% where, $V_a$, $V_w$ are \textcolor{red}{????????????, and ????????}. 
%Here, $V_g = \sqrt{(Va_x+W_x)^2 + (Va_y+W_y)^2}$ and $\chi= \tan^{-1}((Va_y+W_y)/(Va_x+W_x))$, where Va and W denote the UAV airspeed and Wind speed, respectively. And, subscripts $x$ and $y$ denote $X$ and $Y$ directions of inertial reference frame.
Incorporation of ground speed ($V_g$) and course of the UAV ($\chi$) in \eqref{eq:dynamics_chi_dot} ensures that the wind-effect is implicitly factored in \cite{nelson2007}.
%The change of heading will guide the UAV to change its direction. The dynamics are presented as,
The UAV's guidance command is given by the turn rate ($\Dot{\chi}$), where
% \begin{align}
%     \Dot{\chi} = \alpha(\chi^c-\chi)
%     \label{eq:chi_dot}
% \end{align}
%Here, $\alpha$ is a fixed positive value while $\chi^c$ represents the input command given to the UAV. In order to follow a desired path, an appropriate command value for $\chi^c$ needs to be determined. The main objective of this paper is to formulate the $\chi^c$ that allows the UAV to follow a desired path in less time while keeping the rate of change of $\chi$ values within a specified range.
$\alpha$ is a positive constant, % (\textcolor{green}{1/$\mathbf{\alpha}$ is time constant}) 
and $\chi^c$ represents the input course command generated by the guidance law for the UAV to follow a desired general path with time-efficiency, turn rate feasibility and path curvature constraint. % a suitable value of $\chi^c$ needs to be generated by the developed path following guidance.
%%%%%%%%%%%%%%%%%%%%%%%%%%%%%%%%%%%%%%%%%%%%%%%%%%%%%%%%%

\subsection{Definitions Related to Reference Path}
%The path following problem is described as having UAVs follow a predefined reference path in space. 
In this paper, the closest point on the reference path from the UAV's current position $\bm{p}$ is considered as the reference point $(\bm{p}_{ref})$. And, the distance from $\bm{p}$ to $\bm{p}_{ref}$, represents the cross-track error $d=\rho\|\bm{d}\|$,
% And, $d=\rho\|\bm{d}\|$ represents cross-track error i.e. the distance from the UAV’s current position  $\bm{p}$ to $\bm{p}_{ref}$.
where, $\rho=\pm 1$ represents the side of the reference path the UAV is lying on. For example, 
% in the path following geometry shown 
in Fig. \ref{fig:VF_for_line_following}, $\rho=+1$ as the UAV is on right side of the reference path. 
 % In this paper, the reference path $\mathcal{P}_D$ is represented as an ordered set of waypoints ($\bm{p}_D(\gamma) \in \mathbb{R}^2$), where $\gamma(=1,2,\ldots,n)$ are the indices of waypoints.
%
%nn If the reference path is available as a continuous function $\mathcal{P}_D$, it can be sampled at regular intervals and $\mathbb{P}_D$ can be obtained. 
% The reference path to follow is described in this paper as a continuous function of the scalar parameter $\gamma$.
% \begin{align}
% %\begin{array}{l}
%     &\mathbf{{\mathcal{P}_D}(\gamma)} = \{\mathbf{p_D}(\gamma)\}_{\gamma_{start}}^{\gamma_{end}} = [x_d(\gamma)\quad         y_d(\gamma)]^T;\nonumber \\
%     &\mathbf{{p_D}}^{start} \equiv {\mathcal{P}_D}(\gamma_{start});\quad 
%     \mathbf{{p_D}}^{end} \equiv {\mathcal{P}_D}(\gamma_{end}) 
% %\end{array}
% \label{path_definition}
% \end{align}
%\subsection{Closest Point}
%The closest point from the UAV's current position $\bm{p}$ to $\mathbb{P}_D$ is considered as the reference point. Its index is obtained as, 
%\begin{align}
 %   \label{eq:ref_point}
    %\gamma_{\mathrm{ref}}=\arg\min_{\gamma}(||\bm{p}-\bm{p}_D(\gamma)||)
%\end{align}
%The UAV's cross-track error vector w.r.t. the reference path is represented by $\bm{D}\triangleq(\bm{p}-\bm{p}_D(\gamma_{\mathrm{ref}}))$, and the cross-track error as $d=\bm{|D|}$. 
The tangent vector to the path at $\bm{p}_{ref}$ is denoted as $\bm{T}^p$, which is perpendicular to $\bm{d}$. Here,
% Note that $\bm{T}^p$ and $\bm{d}$ vectors are perpendicular to each other. Thus, %The rate of change of the cross-track error $d$ is given as, 
\begin{align}\label{eq:d_dot}
    \Dot{d}= V_g\sin(\chi -\chi^p)
\end{align}
 and, $\chi^p$ is the angle between $\bm{T}^p$ and $X_I$-axis. 
%\begin{align}\label{eq:ref_point_slope}
 %   \chi^p =\tan^{-1}({\Delta y_D(\gamma)}/{\Delta x_D(\gamma)})
%\end{align}
%where, $\Delta y_D(\gamma) = y_D(\gamma)-y_D(\gamma -1) $, and $\Delta x_D(\gamma)= x_D(\gamma)-x_D(\gamma -1) $. 
% \begin{remark}
%     The tangent vector $T^p$ and the d vector will be perpendicular to each other. And the rate of change of d will be 
%     \begin{align}\label{eq:d_dot}
%         \Dot{d}=\rho V_g\sin(\chi -\chi^p)
%     \end{align}
% Where, $\rho=\pm 1$, represents the side of the path UAV is relying on. $\chi$ is the course angle, and $\chi^p$ is the angle between $T^p$ and $X_I$-axis. where, 
% \begin{align}\label{eq:ref_point_slope}
%     \chi^p =\tan^{-1}({\Delta y_d(\gamma)}/{\Delta x_d(\gamma)})
% \end{align}
% \end{remark}
\section{Guidance for General Reference Path Following} \label{sec:line_VF}
%\textcolor{red}{GENERAL INTRODUCTION}
In this section, a guidance strategy is formulated for an UAV to follow a general reference path based on a novel switched vector field-based method leading to enhanced performance in terms of path following time, while also satisfying feasible turn rate and path curvature constraints.

\subsection{Vector Field Description} \label{subsec:line_VF_description}

%%%%%%%%%
%%%%%%%%%
\subsubsection{Dependence of Vector Field on Cross-track Error}
\label{subsubsec:line_VF_based_on_distance}
\begin{figure}[t!]
    \centering
    \begin{subfigure}[b]{0.65\linewidth}        %% or \columnwidth
        \centering
        \includegraphics[scale=0.55]{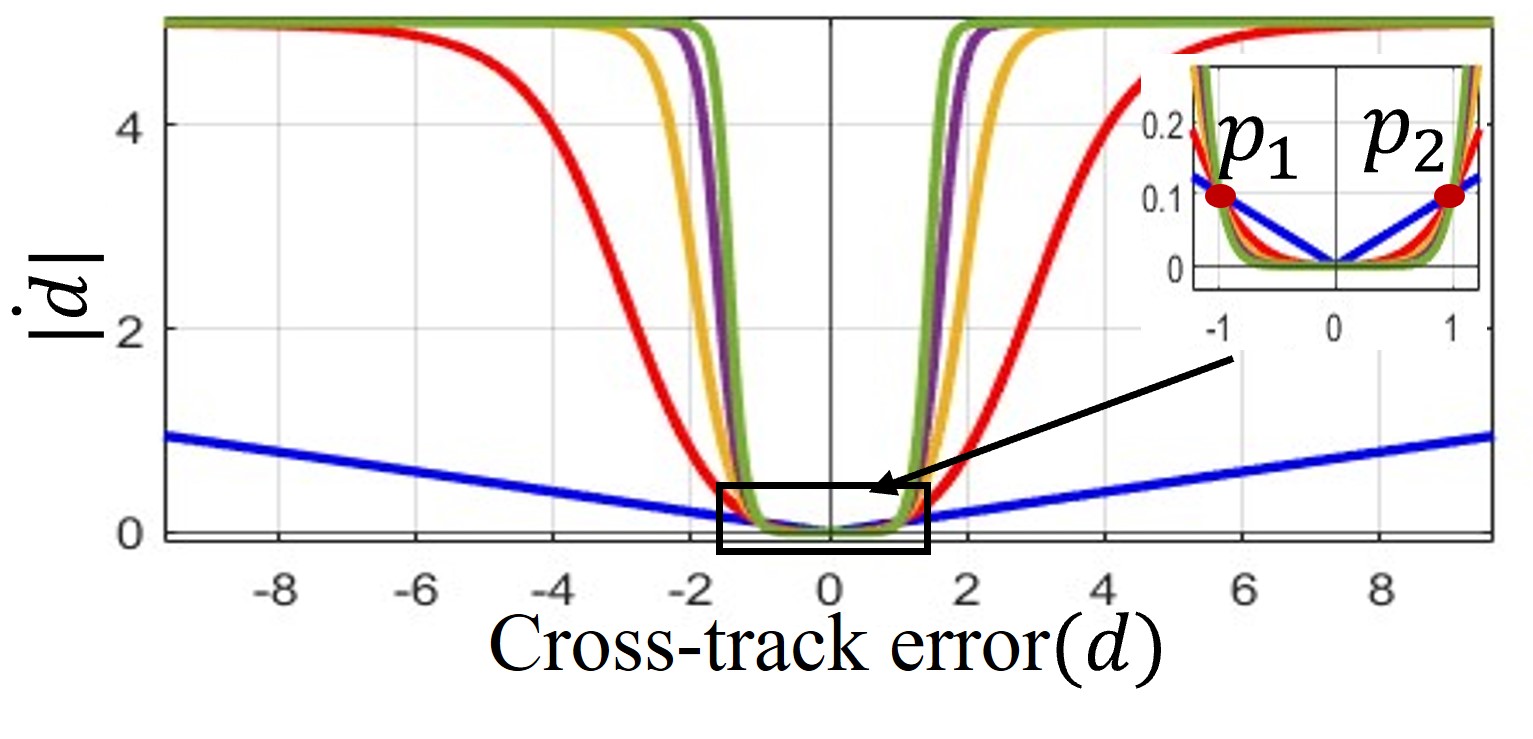}
         \caption{$\vert \Dot{d} \vert$ vs $d$}
         \label{fig:d_dot_comparison}
    \end{subfigure}
    \begin{subfigure}[b]{0.65\linewidth}        %% or \columnwidth
        \centering
        \includegraphics[scale=0.55]{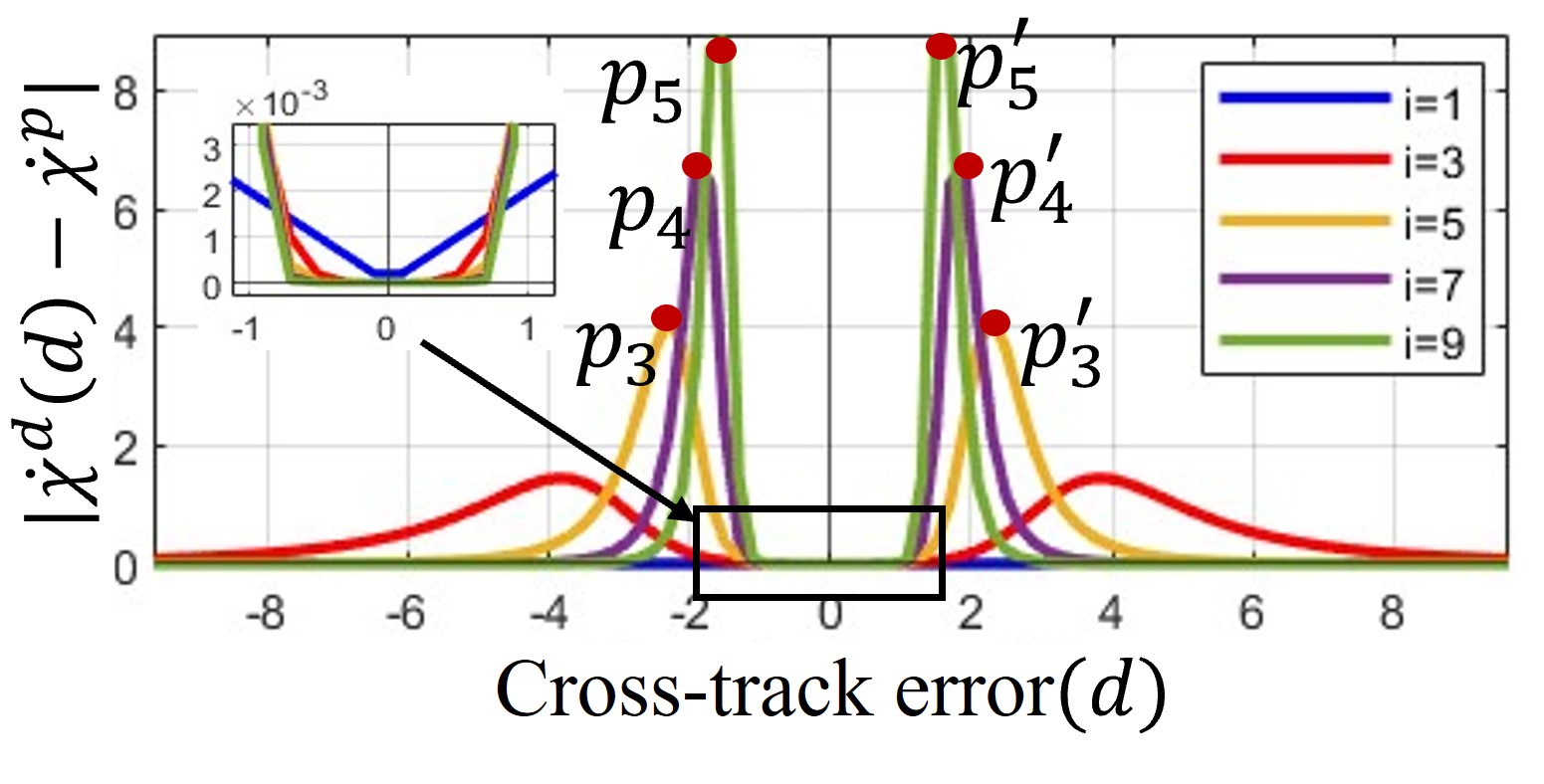}
         \caption{$|\Dot{\chi}^d(d)-\Dot{\chi^p}|$ vs $d$}
         \label{fig:chi_d_dot_comparison}
    \end{subfigure}
    \caption{Variation of $\vert \Dot{d} \vert$ and $|\Dot{\chi}^d(d)-\Dot{\chi^p}|$, for  $i=1,3,5,7,9$.}
    \label{fig:comparison}
\end{figure}

% For positive y direction the $\chi \in (0,\pi)$. As a result, $\Dot{y}>0$, and for negative y direction $\chi \in (0,-\pi)$, so $\Dot{y}<0$.
%The cross-track error of the UAV w.r.t. the reference path is represented by $d$, and the course angle is represented by $\chi$. 
The vector field should be such that when the UAV is far away from the desired path (large $\vert d \vert$), the desired course is directed towards the path with a constant course angle $\chi^p - \chi^\infty$, and as $\vert d \vert$ approaches zero, the desired course aligns with the desired path. To achieve this, for general reference path following, the desired vector field is considered as follows.
\begin{equation}
    \small
    \chi^d(d)=\chi^p -\chi^\infty({2}/{\pi}) \tan^{-1}(k_i d^i)
    \label{eq:chi_d_d^i}
\end{equation}
where, $i=2n+1$ and $n$ is an integer, $k_i$ is a positive constant that affects the rate of transition from $\chi^{\infty}$ to $\chi^p$ while guiding the UAV to reach the desired path asymptotically. Note that the vector field considered in  \cite{nelson2007}, \cite{griffiths2006vector} is a special Case of Eq. \eqref{eq:chi_d_d^i} with $i=1$. When the UAV is able to follow the vector field accurately, that is $\chi = \chi^d(d)$, from \eqref{eq:d_dot}, 
% then the cross-track error variation can be expressed from \eqref{eq:d_dot} as,
\begin{equation*} \label{eq:d_dot_new}
    % \begin{split}
    \small
\frac{d(d)}{dt} =V_g\sin(\chi^d(d) -\chi^p) =-V_g\sin(\chi^\infty\frac{2}{\pi} \tan^{-1}(k_i d^i)).
% \end{split}
\end{equation*}
Note that for higher value of $i$, %when the UAV is far away from the desired path, that is 
when $\vert d \vert$ is large, $\vert\Dot{d}\vert$ is also higher thus enabling the UAV to come closer to the desired path smoothly at a faster rate. But, when %the UAV is very close to the desired path, that is 
$\vert d \vert$ is very small, say $d$ is within the interval $(-d_s,d_s)$ (shown as $(p_1,p_2)$ in Fig. \ref{fig:d_dot_comparison}), 
% the cross-track error convergence rate 
$\vert\Dot{d}\vert$ is highest for $i=1$. Then, cross-track error-dependent switching of vector field could be considered at $d=\rho d_s$ to ensure faster convergence.
%and with increasing value of $i$, $\Dot{d}$ will decrease as illustrated in Fig.~\ref{fig:d_dot_comparison}. 
Now, for $\chi=\chi^d$, 
% the rate of change of desired heading angle,
\begin{flalign*}\label{eq:Chi_d_dot}
\small
    \frac{d\chi^d(d)}{dt}
    % &= \Dot{\chi^p}-\chi^{\infty} \frac{2}{\pi}\frac{i k_i d^{i-1}}{1+(k_i d^i)^2}\Dot{d}\\ \nonumber
    = &\Dot{\chi^p}-\chi^{\infty} \frac{2}{\pi}\frac{i k_i d^{i-1}}{1+(k_i d^i)^2}V_g\sin(\chi^\infty\frac{2}{\pi} \tan^{-1}(k_i d^i)).
\end{flalign*}
 %\textcolor{red}{Here, note that $|\Dot{\chi}^d(d)-\Dot{\chi^p}|$ is higher for higher value of $i$, which may lead to a requirement of very high turn rate of the UAV. For example, as can be seen in Fig.~\ref{fig:chi_d_dot_comparison}, at $p_3$, $p_4$, $p_5$ and $p^{'}_3$, $p^{'}_4$, $p^{'}_5$ points, when $i>3$ the generated turn rate command is very high (even greater than $\pi$ rad/s). Thus, it is evident that although for large cross-track error, higher $i$ leads to faster convergence rate, it also demands higher turn rate command, and for $i>3$ it leads to excessively high, often infeasible, turn rate command.}
Here, as $i$ increases, $|\Dot{\chi}^d(d)-\Dot{\chi^p}|$ also increases for large $|d|$, which in turn demands high turn rate of the UAV. This is illustrated in Fig.~\ref{fig:chi_d_dot_comparison}. For $i>3$, turn rate command becomes  excessively high, %even greater than $>\pi$ rad/s, 
which is often infeasible (see $p_3$, $p_4$, $p_5$ and $p^{'}_3$, $p^{'}_4$, $p^{'}_5$ points in Fig.~\ref{fig:chi_d_dot_comparison}).

% \begin{figure}[]
%     \centering
%     \includegraphics[scale=0.4]{figures/convergence_new.png}
%     \caption{Variation of $\chi^d(d)$ and $\Dot{d}$ with cross-track error (d)}
%     \label{fig:Convergence}
% \end{figure}
% Let's consider $d$ as the distance from the UAV's current position to the desired path. Therefore if the desired path is a straight line aligned with the x-axis, y will represent the distance. So the convergence rate,
% \begin{equation} \label{}
% \begin{split}
% \frac{d}{dt}(d) & = \frac{d}{dt}(y) = \Dot{y} = V_g \sin(\chi) = V_g \sin(\chi^d(d))\\
% \end{split}
% \end{equation}
% So based on the $\chi^d(d)$ function, the rate of change of convergence will vary. So when the distance is higher the switching distance $d_s$,i.e. from the fig \ref{fig:Convergence} (b), when d lies in $d<p_1$ or $d>p_2$ range, the $\Dot{d}$ change will be high for $\chi^d(d)$ as eq. \eqref{eq:chi_d_d^3}. Similarly, when the distance is smaller than the $d_s$,i.e. from the fig \ref{fig:Convergence} (b), when d lies in $P_1<=d<=p_2$, the $\Dot{d}$ change will be high for $\chi^d(d)$ as eq. \eqref{eq:chi_d}. 

Thus, to mitigate these issues while also deriving the advantage of both the formulations of $\chi^d(d)$ with $i=1$ and $i=3$, a switching mechanism is considered in the vector field formulation based on whether $|d|$ is greater or less than $d_s$, which % $d$ enters the close range of reference   path.
% the interval $[p_3,p_4]$. 
ensures the UAV to converge with the desired path smoothly at a faster convergence rate without requiring infeasibly high turn rate. %When the UAV is away from the desired path, it is required for the UAV to reach the desired path with a faster convergence rate, and align with the desired path direction.
Thus, $i=3$ is considered for $|d|\geq d_s$, and $i=1$, else. % and define $d_{s}$ as the cross-track error distance, at which the desired course angle $\chi^d(d)$ switches from $i=3$ to $i=1$ in \eqref{eq:chi_d_d^i}. 
In order to attain a continuous profile of $\chi^d(d)$, the convergence rate controlling parameters $k_1$ and $k_3$ are so chosen that at $d=\rho d_s$, the $\chi^d(d)$ is same for both $i=1$ and $i=3$. This implies that $d_s=\sqrt{{k_1}/{k_3}}$. 
% $d_s$ is obtained as,
% \begin{equation} \label{eq:switching_distance}
% d_s=\sqrt{{k_1}/{k_3}}
% \end{equation}
Thus, the switched vector field based on cross-track error is expressed in terms of the desired UAV course as, %a function of $d$ as below.
\begin{equation}
    \label{eq:chi_d_Switched}
    % \[
\chi^d(d) =
\begin{cases}
\chi^p-\chi^\infty({2}/{\pi}) \tan^{-1}(k_3 d^3) & \text{if $\vert d\vert>d_s$} \\
\chi^p-\chi^\infty({2}/{\pi}) \tan^{-1}(k_1 d) & \text{if $\vert d\vert\leq d_s$} 
\end{cases}
% \]
\end{equation}
Note that following \eqref{eq:chi_d_Switched}, $\chi^d(d)$ approaches to $\chi^p-\chi^\infty$ and $\chi^p$ as $d$ approaches $\infty$ and zero, respectively. %\textcolor{red}{It also ensures faster approach to the reference path for both $\vert d\vert>d_s$ and $\vert d\vert \leq d_s$}.

% \textcolor{red}{ d dot RELATION FOR CURVATURE CONSTRAINTS SECTION}
% \textcolor{blue}{Form eq~\eqref{eq:d_dot}, when $\chi^d(d)$ is expressed as \eqref{eq:chi_d_d^3},
% \begin{equation}
%     \Dot{d}= -\dfrac{k_3V_\text{g}d^3}{\sqrt{k_3^2d^6+1}}
% \end{equation}
% And for \eqref{eq:chi_d}
% \begin{equation}
%     \Dot{d}= -\dfrac{k_1V_\text{g}d}{\sqrt{k_1^2d^2+1}}
% \end{equation}
% Therefore, the convergence rate is also dependent on $k_1$ and $k_3$ values. With higher values of $k_1$ and $k_3$ the convergence rate will increase. However, the maximum value of $k_1$ and $k_3$ is restricted by the maximum allowable turn rate, which will be discussed in section~\ref{sec:Curvature_Constraints}.}

%\textcolor{red}{Alternative-2 ends}
\vspace{5pt}
\subsubsection{Dependence on Instantaneous UAV Course Angle ($\chi$)}
\label{subsubsec:line_VF_based_on_Heading}
\begin{figure}
    \centering
    \includegraphics[scale=0.9]{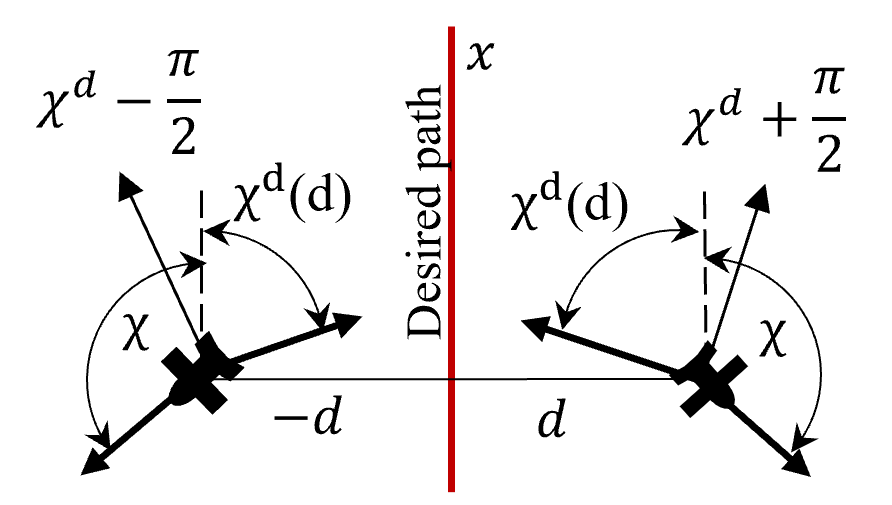}
    \caption{Course angle based vector field for $\chi^p=0$}
    \label{fig:chi-chi_d_High}
\end{figure}

When
% $\chi$ is highly deviated from $\chi^d(d)$ obtained in \eqref{eq:chi_d_Switched}, that is when 
$|\chi-\chi^d(d)|$ is very high, as illustrated in Fig.~\ref{fig:chi-chi_d_High}, %is looking in the opposite direction the difference between the current heading and the desired heading will be very high. If $|\chi-\chi^d(d)| > \pi/2$ 
then the generated guidance command also becomes very high. % leading to a demand of infeasible turn rate command. 
%Illustration of such scenario is shown in Fig.~\ref{fig:chi-chi_d_High}. 
This problem becomes more severe when $\vert d\vert >d_s$ as more aggressive turn rate command is generated by using the formulation of $\chi^d(d)$ with $i=3$ than that with $i=1$. In order to get rid of this problem, 
% the current course of the UAV is
$\chi$ also needs to be incorporated in the formulation of $\chi^d(d)$. Hence, the switched vector field, based on current course angle becomes,
%Assuming $\chi^d(d)$ as the desired heading at some instantaneous distance, the proposed vector field can be represented as,
\begin{equation}\label{eq:chi_d_turn_line}
\small
\chi^d(d,\chi) =
\begin{cases}
\chi^d(d) + \rho \frac{\pi}{2},& \text{if } \vert d \vert >d_s, |\chi -\chi^d(d)|>\frac{\pi}{2}\\
\chi^d(d) & \text{else} 
\end{cases}
\end{equation}
where, $\chi^d(d)$ is as given in \eqref{eq:chi_d_Switched}.
% and $\rho$ represents the side of the reference path the UAV is lying on. 
Note from Fig.~\ref{fig:chi-chi_d_High} that the selection of $\chi^d(d,\chi)$ in \eqref{eq:chi_d_turn_line} above leads to less turn rate command than $\chi^d(d)$ in \eqref{eq:chi_d_Switched} by decreasing $|\chi-\chi^d(d,\chi)|$ for $|\chi -\chi^d(d)|>{\pi}/{2}$. This, in effect, helps in ensuring that the commanded turn rate is within the upper threshold that arises from kino-dynamic constraints of a real unmanned vehicle.
    % \textcolor{red}{Note that following \eqref{eq:chi_d_turn_line}, $\chi^d(d)$ will decrease with increasing d, and will approach $-\chi^\infty$ as d approaches infinity.}
%     and $|\chi-\chi^d(d)| > \pi/2$ ?????
% Note that following \eqref{eq:chi_d_turn_line}, \textcolor{red}{chi-d approaches ???? as d approaches infinity and $|\chi-\chi^d(d)| > \pi/2$ ??????}
% So, when $|\chi-\chi^d(d)| > \pi/2$, the $\chi^d(d) + \rho \frac{\pi}{2}$ will be desired direction for UAV to follow, instead of $\chi^d(d)$, as result, the required turn rate command will be reduced.   
\subsection{Overall Guidance Strategy}
\label{subsec:line_Overall_Guidance_Strategy}
Based on the discussion in Section \ref{subsec:line_VF_description}, the overall structure of the novel switched vector field is presented in terms of desired course angle expressed as a function of both cross-track error and course angle of the UAV as,
% The desired vector field described in Section~\ref{sec:line_VF} can be compactly written as given by 
% \eqref{eq:chi_d_turn_line_final},\eqref{eq:chi_d_qube_final}, and \eqref{eq:chi_d_final}. 
\begin{numcases}
 {\chi^d(d,\chi) =}
 \label{eq:chi_d_turn_line_final}
        \small
         \begin{split}
      \chi^d(d)+\rho\frac{\pi}{2} = \chi^p-\chi^\infty\frac{2}{\pi} \tan^{-1}(k_3 d^3) +\\ \rho{\pi}/{2}, \quad \text{if } \vert d \vert>d_s, |\chi -\chi^d(d)|>{\pi}/{2};        
        \end{split} \\
               \label{eq:chi_d_qube_final} 
        \small
        \begin{split}
              \chi^d(d) =  \chi^p -\chi^\infty\frac{2}{\pi} \tan^{-1}(k_3d^3),\\ \text{if } \vert d \vert> d_s,
                 |\chi -\chi^d(d)| \leq {\pi}/{2}; 
        \end{split} \\
               \label{eq:chi_d_final}
        \small             
         \begin{split}
             \chi^d(d) = \chi^p -\chi^\infty\frac{2}{\pi} \tan^{-1}(k_1 d), \text{if } \vert d \vert \leq d_s.
       \end{split}        
\end{numcases} 
This desired vector field can be followed using the guidance command in \eqref{eq:dynamics_chi_dot}, where the commanded UAV course $\chi^c$ is: 
% \eqref{eq:chi_c_turn_line},\eqref{eq:chi_c_d^3}, and \eqref{eq:chi_c_d}, respectively.
\begin{numcases}{{\chi^c=} }
\label{eq:chi_c_turn_line}
\small
\begin{split}
    \chi+\frac{\Dot{\chi^p}}{\alpha}-\frac{1}{\alpha}\chi^\infty\frac{2}{\pi} \frac{3k_3 d^2}{1+(k_3d^3 )^2}V_g \sin(\chi-\chi^p)\\ -\rho \frac{\eta}{\alpha}(|\Tilde{\chi}|)^{n/m},\quad \text{if } \vert d \vert >d_s, |\chi -\chi^d(d)|>\frac{\pi}{2};
\end{split}\\
         \label{eq:chi_c_d^3}
\small
 \begin{split}
     \chi+\frac{\Dot{\chi^p}}{\alpha} -\frac{1}{\alpha}\chi^\infty\frac{2}{\pi} \frac{3k_3d^2}{1+(k_3d^3)^2}V_g \sin(\chi-\chi^p)\\- \frac{\beta}{\alpha}\sign({\Tilde{\chi}}),\quad  \text{if }\vert d \vert>d_s, |\chi -\chi^d(d)| \leq \frac{\pi}{2};
 \end{split}  \\      
        \label{eq:chi_c_d}
\small        
\begin{split}
     \chi +\frac{\Dot{\chi^p}}{\alpha}-\frac{1}{\alpha}\chi^\infty\frac{2}{\pi} \frac{k_1}{1+(k_1d)^2}V_g \sin(\chi-\chi^p)\\- \frac{\beta}{\alpha}\sign({\Tilde{\chi}}),\quad \text{if } \vert d \vert \leq d_s. 
\end{split}        
 \end{numcases}
%  Here,
% \begin{align*}
% \small
% \sat(\frac{\Tilde{\chi}}{\epsilon}) =
% \begin{Cases}
% \frac{\Tilde{\chi}}{\epsilon},& \text{if } \quad |\frac{\Tilde{\chi}}{\epsilon}|\leq 1\\
% \sign(\frac{\Tilde{\chi}}{\epsilon}), & \text{else} 
% \end{Cases} 
% \end{align*}
% And,
% \begin{align*}
% \sign(\frac{\Tilde{\chi}}{\epsilon}) =
% \begin{Cases}
%  1,& \text{if  } \quad |\frac{\Tilde{\chi}}{\epsilon}| >0\\
%  0,& \text{if  } \quad |\frac{\Tilde{\chi}}{\epsilon}| = 0\\
% -1, & \text{if  } \quad |\frac{\Tilde{\chi}}{\epsilon}| < 0
% \end{Cases}
% \label{}
% \end{align*}
where, $\Tilde{\chi}\triangleq\chi -\chi^d(d,\chi)$. And, m, n are odd and co-prime integers such that $0 <n<m$. And, $\beta = {\sigma}/({1+|\Tilde{\chi}|})$, where $\sigma>0$. % so $\beta>0$ . 
Here, $\beta$ and $\eta$ ($>0$) control the shape of the trajectories onto the sliding surface $\Tilde{\chi}=0$. 
 
% Here, $\sat(x)$ is defined as $\sat(x) =x$ if $|x|\leq 1$, and $\sat(x) =\sign(x)$, else. %Similarly, if $|x| > 0$, $\sign(x)=1$, if $|x| = 0$, $\sign(x)=0$, and if $|x| < 0$, $\sign(x)=-1$.  
%$\epsilon >0$ defines the width of the boundary region around the sliding surface that is used to reduce chattering in the turn rate command.
% \RestyleAlgo{ruled}
% \begin{algorithm}[h!]
% \caption{Switched VF-based Path Following Guidance}\label{alg:two}
% \textbf{Obtain:} Initial UAV position and Heading, details of the reference path.\\ 
% \textbf{Fix guidance parameters:} $\alpha$, $\chi^\infty$, $k_1$, $k_3$, $\sigma$, $\beta$, $\eta$, $n$ \& $m$\\
% % \textbf{Initialize:} write\\
% \textbf{Set}  Switching distance $d_{s}$

% \While{$\gamma_{end}$ $>$ $\gamma_{\mathrm{ref}}$}{
%   \eIf{$d >$ $d_{s}$}{
%     \eIf{$|\chi - \chi^d(d)| > \frac{\pi}{2}$}{
%      Define $\chi^d(d,\chi)$ as  \eqref{eq:chi_d_turn_line_final} \;
%      Calculate $\chi^c$ using \eqref{eq:chi_c_turn_line}\;
    
%   }{
%      Define $\chi^d(d,\chi)$ as  \eqref{eq:chi_d_qube_final} \;
%      % 
%      Calculate $\chi^c$ using \eqref{eq:chi_c_d^3}\;
%   }
%   }{
%     Define $\chi^d(d,\chi)$ as  \eqref{eq:chi_d_final} \;
%      Calculate $\chi^c$ using \eqref{eq:chi_c_d}\;
%   }
%   Generate $\Dot{\chi}$ in \eqref{eq:dynamics_chi_dot} using updated $\chi^c$ \\
%   Update the vehicle position and course\\
% }
% \end{algorithm}

%%%%%%%%%%%%%%%%%%%%%%%%%%%%%%%%%%%%%%%%%%%%%%%%%%%%%%%%
\subsection{ Convergence Analysis 
% of the Developed Guidance Strategy
}
\label{subsec:Line_analysis}
Consider $\chi^\infty=\pi/2$
% For the convergence analysis, first consider 
and the scenario $d>0$. The convergence analysis for the path following engagement geometries mentioned in \eqref{eq:chi_c_turn_line}-\eqref{eq:chi_c_d} is given below.
% To include all path following engagement geometries mentioned in \eqref{eq:chi_c_turn_line}-\eqref{eq:chi_c_d}, consider, initially
% we consider the initial condition to lie in the region satisfying 
% $\vert d \vert >d_s$ and $|\chi -\chi^d(d)|>{\pi}/{2}$, stated as Case-1 below, followed by the analysis for other two Cases. 
The analysis for $d<0$ scenario would follow in similar way.

\subsubsection{Case - 1 $(\vert d \vert>d_s$ \textnormal{and} $|\chi -\chi^d(d)|>({\pi}/{2}))$}\label{sec:Case1_Analysis} In this Case, the desired vector field in terms of $\chi^d(d,\chi)$ is as given in \eqref{eq:chi_d_turn_line_final}, and the commanded course angle $\chi^c$ as mentioned in \eqref{eq:chi_c_turn_line}. 
% This implies that in this Case, initially cross-track error increases.
%Thus, in this Case, the cross-track error increases.
For this geometry, the course error $\Tilde{\chi}=\chi-\chi^d(d,\chi)$ is guaranteed to decrease to zero in finite time as proved below.
% If the heading of the UAV is in the opposite direction of the desired path. As a result, the desired vector field will be \eqref{eq:chi_d_turn_line_final}, and the respective $\chi^c$ command will be \eqref{eq:chi_c_turn_line}. However, because of this at the initial instant, the distance d will increase. 
% \begin{equation}
% \begin{split}
%      \Dot{d}&=V_g sin (\chi)\\
%      &=V_g sin (\frac{\pi}{2}+\chi^d(d,\chi))\\
%      &=V_g \cos(\chi^d(d,\chi))
% \end{split}
%    \label{}
% \end{equation}
% As, $- \frac{\pi}{2}<\chi^\infty\frac{2}{\pi} \tan^{-1}(k_3 d^3)<\frac{\pi}{2}$, the $\Dot{d}$ will be positive. However, in this region, the $\chi$ will decrease.
\vspace{5pt}
\begin{proposition} \label{prop_covergence_of_X_line_phase1}
When $\vert d \vert>d_s$ \textnormal{and} $|\chi -\chi^d(d)|>({\pi}/{2})$, where $\chi^d(d,\chi)=\chi^d(d)+{\pi}/{2}$ as given in \eqref{eq:chi_d_turn_line_final}, applying $\chi^c$ command as given in \eqref{eq:chi_c_turn_line} brings $\Tilde{\chi}$ to zero in finite time.
\end{proposition}
\vspace{5pt}
\begin{proof}\label{prop_covergence_of_X_line_phase1_proof}
%Define $\Tilde{\chi}\triangleq\chi-\chi^d(d,\chi)$.  
% Following $\Dot{\chi}$ from \eqref{eq:dynamics_chi_dot}, 
% % and taking the derivative of $\Tilde{\chi}$ w.r.t. time gives,
% \begin{equation}\small
% \Dot{\Tilde{\chi}}=\Dot{\chi}-\Dot{\chi}^d(d,\chi)=\alpha(\chi^c-\chi)-\Dot{\chi}^d(d,\chi).
% \label{eq:chi_tilde_dot1}
% \end{equation}        
Following $\Dot{\chi}$ from \eqref{eq:dynamics_chi_dot} and considering a Lyapunov candidate function $W_1=\Tilde{\chi}^2/2$.  
\begin{equation}\small 
\label{eq:W_1_dot}
% \begin{split}
\Dot{W_1}  = \Tilde{\chi}\Dot{\Tilde{\chi}} = \Tilde{\chi}(\alpha(\chi^c-\chi) -\Dot{\chi}^d(d,\chi))
% \end{split}
\end{equation}
Considering $\chi^d(d,\chi)$ as in \eqref{eq:chi_d_turn_line_final} for Case - 1, we obtain,
%For $|\chi-\chi^d(d)|\geq\frac{\pi}{2}$ the respective Considering $\chi^d(d,\chi)$, the derivative,
\begin{equation}\small
    \Dot{\chi}^d(d,\chi)= \Dot{\chi^p}-\chi^\infty\frac{2}{\pi} \frac{3k_3d^2}{1+(k_3d^3)^2}V_g \sin(\chi-\chi^p)\label{eq:chi_d_dot_Case1}
\end{equation}
Then, applying $\chi^c$ as in \eqref{eq:chi_c_turn_line}, from \eqref{eq:W_1_dot}, \eqref{eq:chi_d_dot_Case1}, it follows that, in Case-1, where $\Tilde{\chi}=\chi-\chi^d(d,\chi)>0$,
\begin{equation}\small
\Dot{W_1}=-\eta\Tilde{\chi}(\Tilde{\chi})^{n/m} < 0; \quad \Dot{\Tilde{\chi}} = -\eta (\Tilde{\chi})^{n/m}<0
\label{eq:W1_chi-tilde_converge}
\end{equation}
% From \eqref{eq:W1_chi-tilde_converge} above, it is evident that $\Tilde{\chi}$ monotonically decreases to zero \cite{yu1999global}.
Here, for odd and co-prime $n,m$ with $0<n<m$, $\Tilde{\chi}^{n/m}$ has one real solution with same sign as $\Tilde{\chi}$, which implies from \eqref{eq:W1_chi-tilde_converge} that $\Tilde{\chi}$ monotonically decreases and converges to zero in finite time $t_s=({m}/(\eta(m-n)))|\Tilde{\chi}(0)|^{(m-n)/m}$ \cite{yu1999global}. %Moreover, in Case-1 ${\Tilde{\chi}}(0)>\pi/2$, thus,   $\Tilde{\chi}$ converge to zero .
% \begin{equation}\small
%     t_s=({m}/{\eta(m-n)})|\Tilde{\chi}(0)|^{(m-n)/m}
%     \label{eq:t_s}
% \end{equation}
This completes the proof of Proposition \ref{prop_covergence_of_X_line_phase1}.
\end{proof}
\vspace{5pt}
\begin{remark}
\label{reamrk1}
As $\chi^d(d)-\chi^p\in(-\pi/2,0)$ for $\chi^\infty=\pi/2$ (refer to \eqref{eq:chi_d_Switched}), and $\chi -\chi^d(d)>{\pi}/{2}$ in Case - 1 geometry, %that is $\Tilde{\chi}=\chi_0 -\chi^d_0(d)>{\pi}/{2}$, 
% \begin{equation}
    %$\chi_0 -\chi^d_0(d)>{\pi}/{2} \implies (\chi_0 -\chi^p_0)-(\chi^d_0(d)-\chi^p_0)>{\pi}/{2}$,
% \end{equation}
% $\chi_0 -\chi^d_0(d)>{\pi}/{2} \implies (\chi_0 -\chi^p_0)-(\chi^d_0(d)-\chi^p_0)>{\pi}/{2}$,
we have $\chi-\chi^p\in(\pi/2,\pi)$. Then, from \eqref{eq:d_dot}, $\Dot{d}=V_g \sin (\chi-\chi^p)>0$ implying that, $\vert d \vert$ increases with time for this geometry. Even at $t=t_s$, although $\chi(t_s) -\chi^d_{t_s}(d)={\pi}/{2}$, we obtain $\chi(t_s)-\chi^p(t_s)\in(0,\pi/2)$ leading to positive $\Dot{d}(t_s)$. 
%From \eqref{eq:chi_d_dot_Case1}, note that $\Dot{\chi}^d(d,\chi)- \Dot{\chi^p}<0$ as $d$ increases, since $\chi -\chi^p\in(0,\pi)$ in Case - 1. Since $\Tilde{\chi}$ reaches zero by time $t_s$ as guaranteed by Proposition \ref{prop_covergence_of_X_line_phase1}, at time $t_s$, $\chi_{t_s}=\chi^d_{t_s}(d)+\pi/2$, where $\chi^d_{t_s}(d)<\chi^d_0(d)$. Recalling from \eqref{eq:chi_d_Switched} that $\chi^d(d) -\chi^p \in (-\pi/2,0)$ for $d>0$, this implies $\chi_{t_s} -\chi^p \in (0,\pi/2)$ indicating that $\Dot{d}(t_s)=V_g \sin (\chi_{t_s}-\chi^p) > 0$ . Thus, even though $\Tilde{\chi}$ decreases to zero, $\vert d \vert$ keeps on increasing for all time in Case - 1. 
\end{remark}
\vspace{5pt}
% \begin{figure*}[]
%      \centering
%      \begin{subfigure}[b]{0.28\textwidth}
%          \centering
%          \includegraphics[width=\textwidth]{figures/Case7_fig1.png}
%          \subcaption{UAV trajectory}
%          \label{fig:example_trajectory}
%      \end{subfigure}
%      % \hfill
%      \begin{subfigure}[b]{0.28\textwidth}
%          \centering
%          \includegraphics[width=\textwidth]{figures/Case7_fig21.png}
%          \subcaption{Desired vector field and $\Dot{\chi}$}
%          \label{fig:example_chi_d_and_chi_tilde}
%      \end{subfigure}
%      \begin{subfigure}[b]{0.28\textwidth}
%          \centering
%          \includegraphics[width=\textwidth]{figures/Case7_fig31.png}
%          \subcaption{$\Tilde{\chi}$ and distance d}
%          \label{fig:example_d_chi_tilde}
%      \end{subfigure}
%      \caption{Illustration of path following by the proposed switched vector field}
%      \label{fig:example}
% \end{figure*}
\subsubsection{Case - 2 $(\vert d \vert \geq d_s$ \textnormal{and} $|\chi -\chi^d(d)|\leq{\pi}/{2})$}\label{sec:Case2_Analysis}
Proposition \ref{prop_covergence_of_X_line_phase1} guarantees finite time completion of Case-1 geometry, leading to Case-2 geometry, for which $\chi^d(d,\chi)$ and $\chi^c$ are switched to \eqref{eq:chi_d_qube_final} and \eqref{eq:chi_c_d^3}, respectively. % As guaranteed by Proposition \ref{prop_covergence_of_X_line_phase1}, even if the initial geometry belongs to Case - 1, it ends in finite time, and subsequently, this Case begins. As the geometry belongs to Case - 2, the desired vector field in terms of $\chi^d(d,\chi)$ is switched to \eqref{eq:chi_d_qube_final}, and the
% % commanded UAV course angle
% $\chi^c$ is accordingly switched to \eqref{eq:chi_c_d^3}. 
Recall from Remark \ref{reamrk1} that $\Dot{d}(t_s)>0$,  therefore,
% as was found in Section \ref{sec:Case1_Analysis},
$d$ increases at the time of phase transition from Case - 1 to Case - 2. However, as $\Tilde{\chi}=\chi-\chi^d(d,\chi)$ is shown to monotonically decrease in this case, $\chi$ crosses $\chi^p$ at some finite time, and 
$|d|$ also decreases then onward.
% that time onward  $\vert d \vert$ decreases. 
%This $\chi^c$ will reduce the $|\Tilde{\chi}|$ value, as a result the difference between $\chi$ and $\chi^d(d,\chi)$ will reduce. After $|\Tilde{\chi}|\leq \pi/2$ the vector field $\chi^d(d,\chi)$, will switch from eq. \eqref{eq:chi_d_turn_line_final} to \eqref{eq:chi_d_qube_final}. After the initial stage, the distance needs to decrease and should go to $d_s$.
\vspace{5pt}
\begin{proposition} \label{prop_covergence_of_X_Tilde_phase2}
When $\vert d \vert \geq d_s$ \textnormal{and} $|\chi -\chi^d(d)|\leq{\pi}/{2}$, where $\chi^d(d,\chi)=\chi^d(d)$ as in \eqref{eq:chi_d_qube_final}, applying $\chi^c$ command given in \eqref{eq:chi_c_d^3} leads 
% the UAV's course error 
$\Tilde{\chi}$ converge to zero in finite time.
\end{proposition}
\vspace{5pt}
\begin{proof}
Following $\Dot{\chi}$ from \eqref{eq:dynamics_chi_dot} 
% and taking the derivative of $\Tilde{\chi}$ w.r.t. time gives,
% \begin{equation}\small
% \Dot{\Tilde{\chi}}=\Dot{\chi}-\Dot{\chi}^d(d,\chi)=\alpha(\chi^c-\chi)-\Dot{\chi}^d(d)
% \label{eq:chi_tilde_dot2}
% \end{equation}        
and considering a Lyapunov candidate function $W_2=\Tilde{\chi}^2/2$,
\begin{equation}\small 
\label{eq:W_2_dot}
% \begin{split}
\Dot{W_2}  = \Tilde{\chi}\Dot{\Tilde{\chi}} 
  = \Tilde{\chi}(\alpha(\chi^c-\chi) -\Dot{\chi}^d(d))
% \end{split}
\end{equation}
From \eqref{eq:chi_d_qube_final}, for Case - 2 geometry we obtain,
%For $|\chi-\chi^d(d)|\geq\frac{\pi}{2}$ the respective Considering $\chi^d(d,\chi)$, the derivative,
\begin{equation}\small
    \Dot{\chi}^d(d,\chi)=\Dot{\chi^p} -\chi^\infty\frac{2}{\pi} \frac{3k_3d^2}{1+(k_3d^3)^2}V_g \sin(\chi-\chi^p)\label{eq:chi_d_dot_Case2}
\end{equation}
Then, applying the commanded course $\chi^c$ as presented in \eqref{eq:chi_c_d^3}, it follows from \eqref{eq:W_2_dot} and \eqref{eq:chi_d_dot_Case2} that for Case-2,
\begin{equation}
\small
\Dot{W_2}=-\beta\Tilde{\chi}\sign({\Tilde{\chi}}) < 0  %\textcolor{blue}{\Dot{W_2} \leq -\beta | \frac{\Tilde{\chi}}{\epsilon} | 
\implies \Dot{W_2} \leq -{\beta} | \sqrt{W_2} |.
% \quad \textcolor{red}{\Dot{\Tilde{\chi}} = -\beta \sat(\frac{\Tilde{\chi}}{\epsilon})<0}
\label{eq:W1_chi-tilde_converge_Case2}
\end{equation}
% \textcolor{blue}{
% \begin{align}
%    \Dot{W_2} \leq \beta \| \Tilde{\chi} \| \implies \Dot{W_2} \leq \beta \| \sqrt{W_2} \|
%    \label{eq:W1_chi-tilde_converge_Case2_finit_time}
% \end{align}
% }
% From \eqref{eq:W1_chi-tilde_converge_Case2}, 
Therefore, it is evident that $|\Tilde{\chi}|$ monotonically decreases and finally converges to zero in finite time \cite{Sliding_Mode_Control_and_Observation_book}. 
% the $\chi^c$ in \eqref{eq:phase2_chi_c_1st} is updated in \eqref{eq:chi_c_d^3}.
\end{proof}
However, in practice, the $\sign(.)$ function in \eqref{eq:chi_c_d^3} leads to chattering in the control output \cite{Sliding_Mode_Control_and_Observation_book}. To mitigate this, the $\sign(.)$ function can be replaced with $\sat(.)$ function. Therefore, the updated command is given as 
\begin{equation}
    \chi^c= \chi+\frac{\Dot{\chi^p}}{\alpha} -\frac{1}{\alpha}\chi^\infty\frac{2}{\pi} \frac{3k_3d^2}{1+(k_3d^3)^2}V_g \sin(\chi-\chi^p)\\- \frac{\beta}{\alpha}\sat(\frac{\Tilde{\chi}}{\epsilon}),
\end{equation}
where, $\sat(x)$ is defined as $\sat(x) =x$ if $|x|\leq 1$, and $\sat(x) =\sign(x)$, else. And, $\epsilon >0$ sets the boundary region width around the sliding surface to reduce chattering in $\dot{\chi}$.
\vspace{5pt}
\begin{remark}
\label{remark2}
Even if Case-2 is initiated with $\chi-\chi^p \in (0,\pi/2)$, as $\Tilde{\chi}$ monotonically decreases to zero in finite time, as shown in Proposition \ref{prop_covergence_of_X_Tilde_phase2}, and as $\chi^d(d)-\chi^p \in(-\pi/2,0)$ for $\chi^\infty=\pi/2$ (refer to \eqref{eq:chi_d_Switched}), $\chi-\chi^p$ also lies in the interval $(-\pi/2,0)$ as $\chi\to\chi^d(d)$. Besides, as $\Dot{\chi}$ is continuous within the time interval the geometry lies in Case - 2, by intermediate value theorem,  $\chi -\chi^p$ is guaranteed to cross zero in a finite time. And, that time onward, $\Dot{d}=V_g\sin(\chi-\chi^p)<0$ implying monotonically decreasing $\vert d \vert$.
\end{remark}
\vspace{5pt}
% \textit{Remark 2:} Since $\Tilde{\chi}$ monotonically decreases, as shown in Proposition \ref{prop_covergence_of_X_Tilde_phase2}, and  $\chi^d(d)\in(-\pi/2,0)$ for $\chi^\infty=\pi/2$ (refer to \eqref{eq:chi_d_Switched}), that is $\chi^d(d)$ does not become greater than zero, and $\chi -\chi^d(d)\leq{\pi}/{2})$ in Case - 2, $\chi$ is guaranteed to cross zero and fall in the interval $(-\pi/2,0)$ in some finite time. And, that time onward, $\Dot{d}=V_g\sin{\chi}<0$, that is the cross-track error starts monotonically decreasing. 
% \textcolor{blue}{
\begin{remark} \label{remark:chi_chattering}
From \eqref{eq:W1_chi-tilde_converge}, $\dot{\Tilde{\chi}}=0$ at $t=t_s$ as $\Tilde{\chi}(t_s)=0$ for $\Tilde{\chi}=\chi-\chi^d(d,\chi)$ in Case - 1. Hence, if the switching of vector field at the transition from Case - 1 to Case - 2 takes place at $t=t_s$, there is a possibility of undesirable chattering at the transition. To avoid that, this switching may be done at $t=t_s-\epsilon_t$ for some small positive $\epsilon_t$, when $\vert \chi -\chi^d \vert = (\pi/2) +\Delta$ for some small positive $\Delta$. Note that since $\Dot{d}$ and $\Dot{\Tilde{\chi}}$ remain positive and negative, respectively, both at $t=t_s-\epsilon_t$ (follows Case - 1) and at the beginning of Case - 2 geometry after the transition, the switching at $t=t_s-\epsilon_t$ would maintain the same convergence pattern, while also ensuring no undesired chattering during the switching of vector field between Case -1 and Case - 2 geometries.
% % \textcolor{red}
% {From Proposition 1 $\dot{\Tilde{\chi}}<0$, thus $\Tilde{\chi} \rightarrow 0$ in finite time $t_s$. At $t=t_s$, Case-2 begins with $\Dot{\Tilde{\chi}}<0$ as per Proposition \ref{prop_covergence_of_X_Tilde_phase2}. Therefore, theoretically as $\dot{\Tilde{\chi}}<0$ before and after the transition, there is no possibility of chattering. In practical application, to avoid switching closer to $\Tilde{\chi} = 0 $, switching at $t<t_s$ which leads to $\vert \chi -\chi^d \vert = (\pi/2) +\Delta$. As Proposition 2 will also make sure $\Dot{\Tilde{\chi}}<0$, even for $\vert \chi -\chi^d \vert \leq (\pi/2) +\Delta$. Therefor, for all time $\Dot{\Tilde{\chi}}<0$, prevent chattering between Case-1 and Case-2.}
\end{remark}
\vspace{5pt}
\begin{proposition}\label{prop_covergence_of_d_for_phase_2}
When $ \vert d \vert \geq d_s$ \textnormal{and} $|\chi -\chi^d(d)|\leq{\pi}/{2}$, where $\chi^d(d,\chi)=\chi^d(d)$ as given in \eqref{eq:chi_d_qube_final}, applying $\chi^c$ as given in \eqref{eq:chi_c_d^3} as $\chi$ converges to $\chi^d(d)$, 
% the cross-track error 
$|d|$ is guaranteed to monotonically decrease and reach $d_s$ in finite time.  
\end{proposition}
\vspace{5pt}
\begin{proof} \label{proof:prop_covergence_of_d_for_phase_2}
% As $\chi^{\infty}$ is restricted to be in the range $\mathbf{\chi^\infty \in (0,\frac{\pi}{2}]},$ then clearly $- \frac{\pi}{2}<\chi^\infty\frac{2}{\pi} \tan^{-1}(rd^3)<\frac{\pi}{2}$. 
% Also for the respective $\chi^d(d,\chi)$, the derivative,
% \begin{equation}
%     \Dot{\chi}^d(d,\chi)= -\chi^\infty\frac{2}{\pi} \frac{3k_3d^2}{1+(k_3d^3)^2}V_g \sin(\chi)
% \end{equation}
% So, from the definition of $W_1$ for given $\chi^c$ \eqref{eq:chi_c_d^3}, the $\Dot{W_1}$ will be less than zero. As a result, The  $\Tilde{\chi}  \rightarrow 0 \equiv \chi - \chi^d (d,\chi)  \rightarrow 0$ in finite time \cite{khalil2015nonlinear}, i.e. $\chi \rightarrow \chi^d(d)$.
Define a Lyapunov candidate function $W_3=d^2/2$. As $\chi$ converges to $\chi^d(d)$ (by Proposition \ref{prop_covergence_of_X_Tilde_phase2}), from \eqref{eq:d_dot}, 
% and taking the derivative
\begin{equation} \label{eq:W_3_dot}
\small
\Dot{W_3} =d\Dot{d}= -V_gd\sin(\chi^\infty({2}/{\pi}) \tan^{-1}(k_3d^3))
\end{equation}
Here, $\Dot{W_3}<0$, and $\dot{d}\leq -V_g\sin(\chi^\infty\frac{2}{\pi} \tan^{-1}(k_3d_s^3))<0$ for $\vert d \vert>d_s\neq 0$. This leads to finite time convergence of $|d|$ to $d_s$. Thus, the proposition holds. %for Case - 2.
%So, $\Dot{W_1}$ is less than zero for $d\neq 0$. As $\chi = \chi^d(d,\chi)$ the distance $d \rightarrow 0$ asymptotically. As, $d_s>0$, therefore $d$ will reach to $d_s$.
\end{proof}
\vspace{5pt}
\subsubsection{Case - 3 $(\vert d \vert < d_s$)}\label{sec:Case3_Analysis}
Propositions \ref{prop_covergence_of_X_line_phase1}, \ref{prop_covergence_of_X_Tilde_phase2} and \ref{prop_covergence_of_d_for_phase_2}, guaranteed the path following guidance switches to this Case within finite time. Consequently, $\chi^d(d,\chi)$ and $\chi^c$ are switched to \eqref{eq:chi_d_final} and \eqref{eq:chi_c_d}, respectively.
% As guaranteed by Propositions \ref{prop_covergence_of_X_line_phase1}, \ref{prop_covergence_of_X_Tilde_phase2} and \ref{prop_covergence_of_d_for_phase_2}, the path following guidance geometry enters this Case in finite time. Accordingly, the desired vector field in terms of $\chi^d(d,\chi)$ is switched to \eqref{eq:chi_d_final}, and the commanded UAV course angle $\chi^c$ is accordingly switched to \eqref{eq:chi_c_d}. 
Following the same steps as in Case - 2 in Section \ref{sec:Case2_Analysis}, the following result can be shown.
\vspace{5pt}
\begin{proposition} \label{prop_covergence_of_X_Tilde_d_phase3}
When $\vert d \vert < d_s$,
% \textnormal{and} $|\chi -\chi^d(d)|\leq{\pi}/{2}$, 
where $\chi^d(d,\chi)=\chi^d(d)$ as given in \eqref{eq:chi_d_final}, applying $\chi^c$ command as given in \eqref{eq:chi_c_d} leads 
% the UAV's course error 
$\Tilde{\chi}=\chi-\chi^d(d,\chi)$ converge to zero in finite time, and as a consequence, $d$ converges to zero asymptotically. 
\end{proposition}
% \textcolor{blue}{
\vspace{5pt}
\begin{remark}\label{remark:d_chattering}
The transition from Case - 2 to Case -3 takes place at $d=d_s\neq 0$. Also, both at the end of Case - 2 and at the beginning of Case -3 geometries, both $\Dot{d}$ and $\Dot{\Tilde{\chi}}$ are negative. These eliminate the possibility of chattering and any degradation of convergence performance during the switching of vector field from Case -2 to Case - 3. %when From Propositions 2, 3 and 4,  $\Dot{\Tilde{\chi}}< 0$ and also $\Dot{d}<0$ during Case-2 and Case-3 $(\textnormal{for } d \neq 0)$. Since the transition occurs at $\vert d \vert=d_s$, the possibility of chattering during the transition from Case-2 to Case-3 is eliminated.
 % $\vert d \vert \neq 0$ in both Cases and the transition between them depends on $\vert d \vert$, the possibility of chattering is eliminated.
% During Case-2 and Case-3, with \eqref{eq:d_dot} and $\chi^c$ command from \eqref{eq:chi_c_d^3} and \eqref{eq:chi_c_d}, $\Dot{d}<0$ ($\because d \neq 0$). Therefore, at $d\approx d_s$, $\Dot{d}\ngtr 0$, eliminating the possibility of chattering during the transition from Case-2 to Case-3.
% During Case-2 and Case-3, using \eqref{eq:d_dot} and applying $\chi^c$ command as given in \eqref{eq:chi_c_d^3} and \eqref{eq:chi_c_d},  $\Dot{d}<0$ ($\because d \neq 0$ ), thus, at $d\approx d_s$, $\Dot{d}\ngtr 0$ under any circumstance, thus the possibility of chattering at the transition of Case-2 and Case-3 is eliminated.
\end{remark}
\vspace{5pt}
\subsubsection{Final Result}\label{sec:Overall_Analysis}
\begin{theorem} \label{Treorem:SL_proof}
Starting from any $\chi$ and $d$, the overall vector field given by \eqref{eq:chi_d_turn_line_final},\eqref{eq:chi_d_qube_final}, \eqref{eq:chi_d_final}, and the respective guidance command represented by \eqref{eq:chi_c_turn_line},\eqref{eq:chi_c_d^3}, and \eqref{eq:chi_c_d}, guarantee that the UAV is able to follow the desired general reference path asymptotically and reaches its close vicinity in finite time. 
\end{theorem}
\vspace{5pt}
\begin{proof}
From Propositions \ref{prop_covergence_of_X_line_phase1} - \ref{prop_covergence_of_X_Tilde_d_phase3} and selecting $d_s=\sqrt{{k_1}/{k_3}}$ suitably by tuning control parameters $k_1$ and $k_3$ in \eqref{eq:chi_d_Switched}, this result follows in a straight-forward way. 
\end{proof}

\subsection{Curvature Constraints}
\label{sec:Curvature_Constraints}
% \begin{figure}[b!]
%     \centering
%     \includegraphics[width=6cm,height=3cm]{figures/Picture3.png}
%     \caption{Curvature constraints for given path }
%     \label{fig:curvature_constraints}
% \end{figure}
Convergence analysis in Section \ref{subsec:Line_analysis} shows that 
convergence rate of $\chi$ to $\chi^d(d,\chi)$ depends on  $\eta$, $\beta$ and $k_i$ (for $i=1,3$), as indicated by  \eqref{eq:W1_chi-tilde_converge}, \eqref{eq:W1_chi-tilde_converge_Case2} and \eqref{eq:W_3_dot}, respectively. These parameters also affect the curvature of the UAV's path as may be noted from \eqref{eq:dynamics_chi_dot}, \eqref{eq:chi_c_turn_line}-\eqref{eq:chi_c_d}. Accordingly, this section outlines conditions on these parameters to ensure kino-dynamic feasibility of the UAV's commanded path.
% From the convergence analysis results in Section \ref{subsec:Line_analysis}, it may be noted that the convergence of $\chi$ to $\chi^d(d,\chi)$ has their dependence on the guidance parameters $k_i$, $\eta$ and $\beta$. Correspondingly, the curvature constraint of the path traversed by the UAV during the path following has also dependence on them. Considering this, this section provides the condition on those parameters in order to satisfy kino-dynamic feasibility of the UAV's commanded path.
For that, considering $\chi = \chi^d(d,\chi)$, the curvature variation is:
\begin{equation}\small
    \kappa={d\chi}/{ds}={d\chi^d(d,\chi)}/{ds}={(d\chi^d(d,\chi)/dt)}/{(ds/dt)}.
    \label{eq:kappa}
\end{equation}
% Recall that $i=1$ 
For $\vert d \vert \leq d_s$, using $\chi^d(d,\chi)$ from \eqref{eq:chi_d_final} and $\chi^{\infty}={\pi}/{2}$,
% \begin{equation}\label{eq:chi_d_for_curvature}
% \small
%     {d\chi^d(d,\chi)}/{dt}={d\chi^p}/{dt} + {d}/{dt}(-\chi^\infty({2}/{\pi})\tan^{-1}(k_1d))
% \end{equation}
% Now, using $\chi^{\infty}={\pi}/{2}$,
\begin{equation}\label{eq:chi_d_for_curvature}
    \small 
    {d\chi^d(d,\chi)}/{dt}-\Dot{\chi^p}
    % &=\frac{k_1\Dot{d}}{1+(k_1d)^2}\\ \nonumber
    % &=-\chi^\infty\frac{2}{\pi}\frac{k_1 V_g \sin(\chi-\chi^P)}{1+(k_1d)^2}\\ \nonumber
% \end{equation}
% % Now, as $\Tilde{\chi}\rightarrow 0, \implies \chi \rightarrow \chi^d(d,\chi)$
% \begin{flalign}
% \small
    % \frac{d\chi^d(d,\chi)}{dt} 
    % &=-\chi^\infty\frac{2}{\pi}\frac{k_1 V_g \sin(\chi^d(d,\chi)-\chi^P)}{1+(k_1d)^2}\\ \nonumber
    % &=-\frac{k_1 V_g \sin(-\tan^{-1}(k_1d))}{1+(k_1d)^2}\\ \nonumber
    ={k_1^2V_gd}/{\left(1+(k_1d)^2\right)^{{3}/{2}}}
\end{equation}
Note that $\Dot{\chi^p}$ does not have dependence on $\vert d\vert$. Thus, the condition for maximum value of $\Dot{\chi^d}(d,\chi)$ is:
\begin{equation}\label{eq:chi_d_dot_and_chi_p_dot}
\small
\nabla_{d}\Dot{\chi}^d(d)={k_1^2V_g\cdot\left(1-2k_1^2d^2\right)}/{\left(k_1^2d^2+1\right)^{{5}/{2}}}=0
\end{equation}
From \eqref{eq:chi_d_dot_and_chi_p_dot}, for $\vert d \vert \leq d_s$, we obtain that $|\Dot{\chi^d}(d,\chi)-\Dot{\chi^p}|_{max}= ({2}/{3\sqrt{3}})k_1V_g$ at $|d|={1}/({\sqrt{2}\,k_1})$. %\leq d_s$. Thus,
% \begin{equation}\label{eq:k_1_range}
%     \small
%     k_1>1/(\sqrt{2}d_s)
% \end{equation}
% \begin{equation}\small
% |\Dot{\chi^d}(d,\chi)-\Dot{\chi^p}|_{max}= ({2}/{3\sqrt{3}})k_1V_g
% \end{equation}
Similarly, recall that for $\vert d \vert > d_s$, $i=3$, and $\chi^d(d,\chi)$ is given in \eqref{eq:chi_d_turn_line_final} and \eqref{eq:chi_d_qube_final}. Then, at $|d|= {\sqrt[6]{5}}/({\sqrt[3]{2}\sqrt[3]{k_3}})$, $|\Dot{\chi^d}(d,\chi)-\Dot{\chi^p}|_{max}= ({2^{{4}/{3}}{\cdot}5^{{5}/{6}}V_\text{g}\sqrt[3]{k_3}})/{9}$. %$>d_s$. Thus,  
% \begin{equation}\label{eq:k_3_range}
%     \small
%     k_3 < \sqrt{5}/(2d^3_s)
% \end{equation}
% \begin{equation}\small
% |\Dot{\chi^d}(d,\chi)-\Dot{\chi^p}|_{max}= {2^{{4}/{3}}{\cdot}5^{{5}/{6}}V_\text{g}\sqrt[3]{k_3}}/{9}
% \end{equation}
% Thus, \eqref{eq:k_1_range} and \eqref{eq:k_3_range} provide the criteria for selection of $k_1$, $k_3$ and $d_s$. 
% \begin{equation}
%     d_s>1/\sqrt{5}.
% \end{equation}
Now, for $\chi=\chi^d(d,\chi)$, from \eqref{eq:kappa},
\begin{equation}
    \small
    ({|\Dot{\chi^d}(d,\chi)|_{max}})/{V_g} = \kappa_{max}
    \label{eq:kappa_max}   
\end{equation}
where, $\kappa_{max}$ is the maximum allowable curvature in the UAV's path. Then, from \eqref{eq:kappa_max},
\begin{align}
    \small 
    % {|\Dot{\chi^d}(d,\chi)-\Dot{\chi^p}|_{max}} \leq {|\Dot{\chi^d}(d,\chi)|_{max}}+{|\Dot{\chi^p}|_{max}}\\
     {|\Dot{\chi^d}(d,\chi)-\Dot{\chi^p}|_{max}} \leq {|\Dot{\chi^p}|_{max}} + {|\Dot{\chi^d}(d,\chi)|_{max}}\nonumber\\
     \implies {|\Dot{\chi^d}(d,\chi)-\Dot{\chi^p}|_{max}} -{|\Dot{\chi^p}|_{max}} \leq V_g \kappa_{max}.
    \label{eq:chi_d_dot_range}
\end{align}
% Now, from \eqref{eq:kappa_max} and \eqref{eq:chi_d_dot_range}, 
% \begin{equation} \label{eq:kappa_max_3}
%     \small
%      {|\Dot{\chi^d}(d,\chi)-\Dot{\chi^p}|_{max}} -{|\Dot{\chi^p}|_{max}} \leq V_g \kappa_{max}.
% \end{equation}
As, $|\Dot{\chi^d}(d,\chi)-\Dot{\chi^p}|_{max}$ dependent on the $V_g$, $k_1$ and $k_3$ as discussed above, \eqref{eq:chi_d_dot_range} can be rewritten as,
% \begin{equation}\label{eq:kappa_max_final}
%     \small
%     \max \left \{\frac{2}{3\sqrt{3}}k_1,\dfrac{2^\frac{4}{3}{\cdot}5^\frac{5}{6}\sqrt[3]{k_3}}{9} \right\}-\frac{|\Dot{\chi^p}|_{max}}{V_g} \leq \kappa_{max} 
% \end{equation}
\begin{equation}\label{eq:kappa_max_final}
    \small
    \max \left \{{2k_1}/({3\sqrt{3}}),({2^\frac{4}{3}{\cdot}5^\frac{5}{6}\sqrt[3]{k_3}})/{9} \right\}-{|\Dot{\chi^p}|_{max}}/{V_g} \leq \kappa_{max} 
\end{equation}
Thus, for given $\kappa_{max}$ (UAV's turn rate constraint), $|\Dot{\chi^p}|_{max}$ (maximum path curvature) and $V_g$ (UAV speed), the overall switched vector field (\eqref{eq:chi_d_turn_line_final}-\eqref{eq:chi_d_final}) can be designed by suitably selecting $k_1$ and $k_3$ that satisfies \eqref{eq:kappa_max_final}. This can be followed by the commanded course angles (\eqref{eq:chi_c_turn_line}-\eqref{eq:chi_c_d}) with suitable selection of $\eta$ and $\beta$ satisfying the $\kappa_{max}$ constraint for effectively following a general reference path without any undesirable chattering at phase transitions, as explained in Theorem \ref{Treorem:SL_proof}, Remarks \ref{remark:chi_chattering} and \ref{remark:d_chattering}.
%  In Fig~\ref{fig:curvature_constraints}, the $\Dot{\chi^d}(d,\chi)$ variations for different values of $k_1$ with different values of $\Dot{\chi^p}$ is shown. Therefore, when  $\Dot{\chi^p} \neq 0$ for higher $k_1$ values, the $V_g\kappa_{max}$ may overshoot. So based on \eqref{eq:kappa_range_1} the range of $k_1$ and $k_3$ values can be computed. %Here, the considered $\Dot{\chi^p}$ is constant for representation, however for general reference path the $\Dot{\chi^p}$ will vary at every reference point.
% Thus, when $\chi \neq \chi^d(d,\chi)$, following \eqref{eq:chi_c_turn_line}-\eqref{eq:chi_c_d}, the max possible tight turn computed and restricted by,
% \begin{equation}\small
%     {|\Dot{\chi^d}(d,\chi)|_{max}+|\Dot{\chi^p}|_{max}}+K \leq V_g\kappa_{max}
% \end{equation}
% where, $K=\eta (\pi/2)^{n/m}$ for $\vert d \vert>d_s$, $|\chi -\chi^d(d)|>{\pi}/{2}$, while $K=\beta$ in other Cases. Again, when $d$ is almost zero, $(-\Dot{\chi^p}_{max}+\Dot{\chi^d}(d,\chi))_{max}=0$. So, an active velocity control algorithm will allow the UAV to take sharp turns within $\kappa_{max}$ curvature constraint i.e. higher value of $(\Dot{\chi^p})_{max}$ when UAV is on the reference path.

%%%%%%%%%%%%%%%%%%%%%%%%%%%%%%%%%%%%%%%%%%%%%%%%%%%%%%%%%
\section{Simulation Results}\label{sec:Simulation_results}
% \begin{figure*}[]
%      \centering
%      \begin{subfigure}[b]{0.28\textwidth}
%          \centering
%          \includegraphics[width=1.1\textwidth]{figures/PVF_trajectory.jpg}
%          \subcaption{UAV trajectory}
%          \label{fig:example_trajectory}
%      \end{subfigure}
%      % \hfill
%      \begin{subfigure}[b]{0.28\textwidth}
%          \centering
%          \includegraphics[width=0.9\textwidth]{figures/fig_5_2.jpg}
%          \subcaption{Desired vector field and $\Dot{\chi}$}
%          \label{fig:example_chi_d_and_chi_tilde}
%      \end{subfigure}
%      \begin{subfigure}[b]{0.28\textwidth}
%          \centering
%          \includegraphics[width=0.95\textwidth]{figures/fig_5_3.jpg}
%          \subcaption{$\Tilde{\chi}$ and distance d}
%          \label{fig:example_d_chi_tilde}
%      \end{subfigure}
%      \caption{Illustration of path following by the proposed switched vector field}
%      \label{fig:example}
% \end{figure*}
\begin{figure}[]
    \centering
    \begin{subfigure}[b]{0.9\linewidth}        %% or \columnwidth
        \centering
        \includegraphics[width=0.95\textwidth]{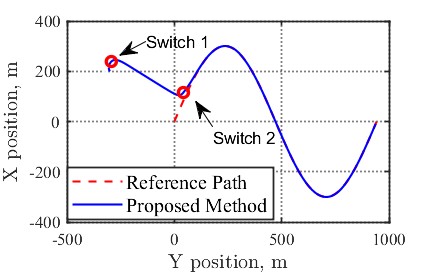}
        \caption{UAV trajectory}
        \label{fig:example_trajectory}
    \end{subfigure}
    \begin{subfigure}[b]{0.9\linewidth}        %% or \columnwidth
        \centering
        \includegraphics[width=0.95\textwidth]{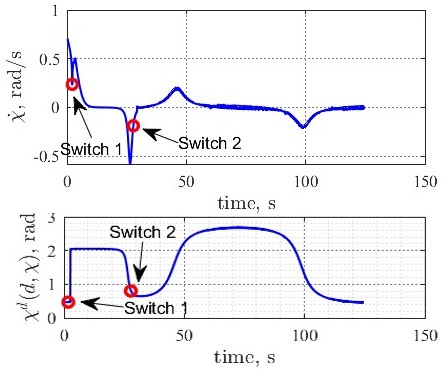}
        \caption{Desired vector field and $\Dot{\chi}$}
        \label{fig:example_chi_d_and_chi_tilde}
    \end{subfigure}
    \begin{subfigure}[b]{0.9\linewidth}        %% or \columnwidth
        \centering
        \includegraphics[width=0.98\textwidth]{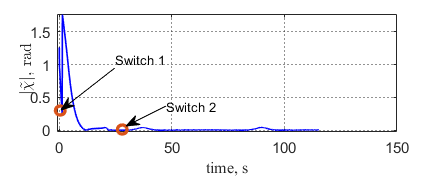}
        \caption{$\vert \Tilde{\chi} \vert $ profile}
        \label{fig:example_chi_tilde}
    \end{subfigure}
    \begin{subfigure}[b]{0.9\linewidth}        %% or \columnwidth
        \centering
        \includegraphics[width=0.9\textwidth]{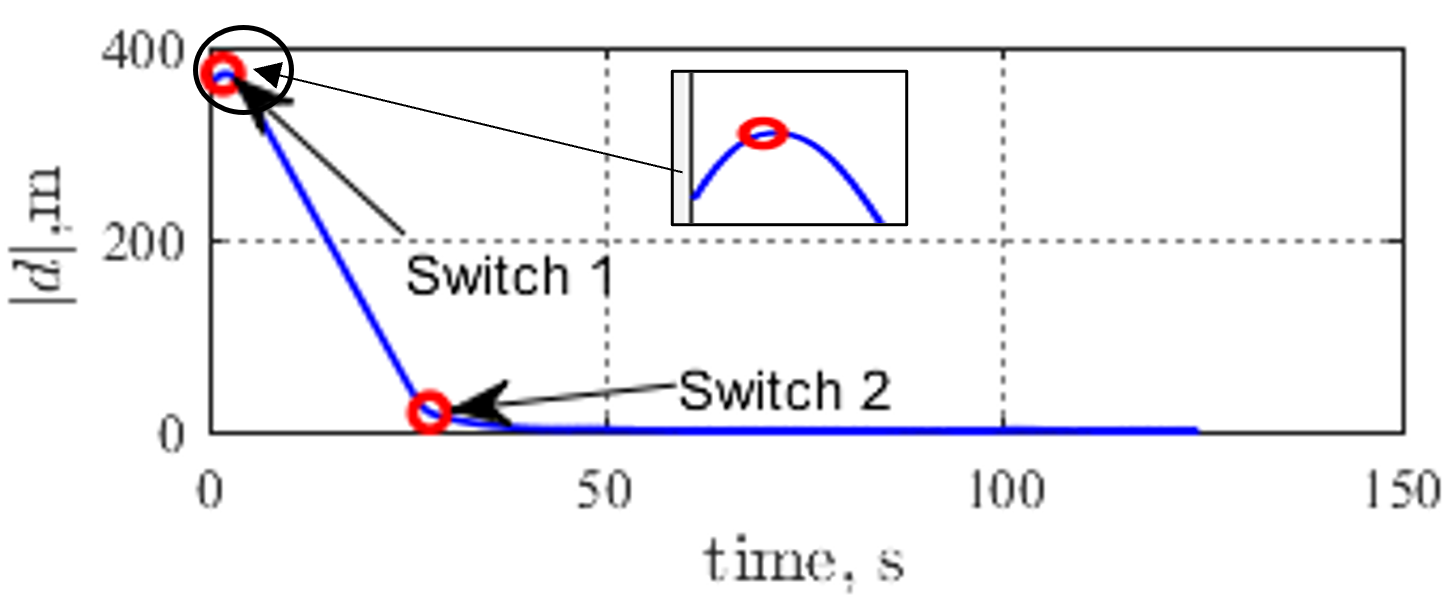}
        \caption{Cross-track error magnitude}
        \label{fig:example_d_chi_tilde}
    \end{subfigure}
    \caption{Illustration of path following by the proposed switched vector field-based guidance.}
    \label{fig:example}
\end{figure}
In this section, the path following guidance strategy presented in Section \ref{subsec:line_Overall_Guidance_Strategy} 
% and \ref{subsec:Circle_Overall_Guidance_Strategy} 
is validated using numerical simulations performed in MATLAB (R2023a) environment. The parameters considered for the simulations are as follow. Constant UAV speed ($V_a$) of  $15$ m/s is considered for all scenarios. And, $\chi^\infty = \pi/2$, $\alpha=1.65$~s$^{-1}$, $\gamma=0.8$, $\eta = \pi/4$~rad, $n=3$, $m=5$, $d_s=10$~m, $k_1=0.01$,  $k_3=k_1/{d_s}^2=10^{-4}$.
% (satisfying \eqref{eq:k_1_range},\eqref{eq:k_3_range}). 
In practice, $\Dot{\chi^p}$ can be approximated using tangents to the path as $\Dot{\chi^p} \approx (\chi^p - \chi^p_{old})/dt$, where $\chi^p_{old}$ is the tangent to the path in the previous time step.
% $k_3$ is calculated using \eqref{eq:switching_distance}. 
% \begin{figure}[]
%      \centering
%      \begin{subfigure}[b]{0.3\textwidth}
%          \centering
%          \includegraphics[width=5cm,height=2.5cm]{figures/comp_PVF_VF_PLOS_L1_trajectory.jpg}
%          \caption{UAV trajectories}
%          \label{fig:Case4_trajectory}
%      \end{subfigure}
%      % \hfill
%      \begin{subfigure}[b]{0.3\textwidth}
%          \centering
%          \includegraphics[width=5cm,height=2.5cm]{figures/comp_PVF_VF_PLOS_L1_chi_dot.jpg}
%          \caption{Turn rate command}
%          \label{fig:Case4_fig2}
%      \end{subfigure}
%      \caption{Illustration of path following by the proposed switched vector field, vector field\cite{nelson2007}, PLOS\cite{kothari2014uav} and $L_1$\cite{park2007}}
%      \label{fig:Case4}
% \end{figure}
\begin{figure}[]
    \centering
    \begin{subfigure}[b]{0.9\linewidth}        %% or \columnwidth
        \centering
        \includegraphics[width=0.9\textwidth]{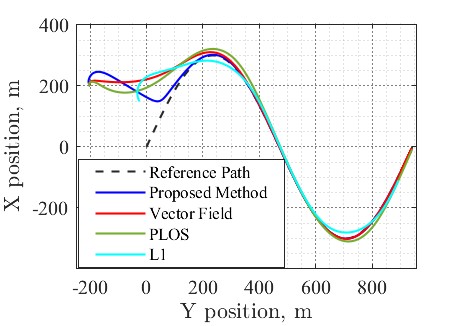}
        \caption{UAV trajectories}
        \label{fig:comparison_trajectories}
    \end{subfigure}
    \begin{subfigure}[b]{0.9\linewidth}        %% or \columnwidth
        \centering
        \includegraphics[width=0.9\textwidth]{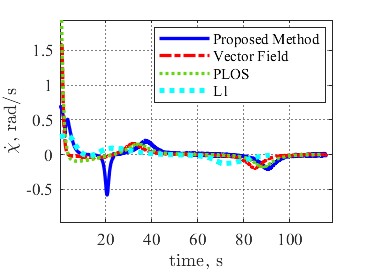}
        \caption{Turn rate command}
        \label{fig:comparison_turn_rate}
    \end{subfigure}
    \caption{Illustration of path following by the proposed method, vector field\cite{nelson2007}, PLOS\cite{kothari2014uav} and NLGL ($L_1$) \cite{park2007}}
    \label{fig:comparison_fig}
\end{figure}
\begin{figure}[]
    \centering
    \includegraphics[scale=0.85]{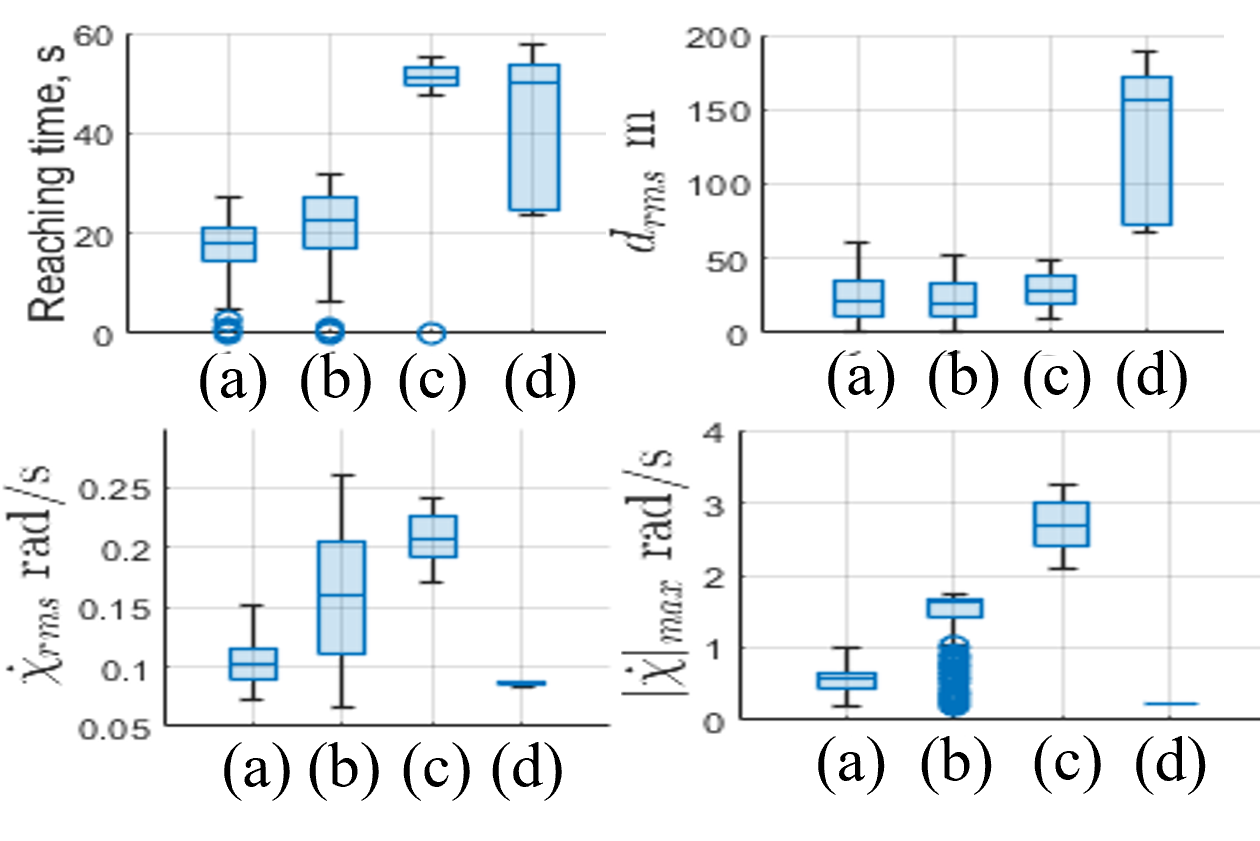}
    \caption{Comparison of path following by the (a) proposed switched vector field, (b) vector field\cite{nelson2007}, (c) PLOS\cite{kothari2014uav} and (d) NLGL($L_1$) \cite{park2007} under random initial condition by Box-plot representation of $t_{conv}$, $d_{rms}$, $\Dot{\chi}_{rms}$, and max($|\Dot{\chi}|$).}
    \label{fig:boxplot}
\end{figure}

Fig.~\ref{fig:example} illustrates a numerical example of following a reference sinusoidal path ($y=r\sin(x)$, where $r=300$ and $x \in [0,2\pi]$) with no wind consideration, that is $V_g=V_a$, for better understanding of the convergence analysis given above. As can be seen in Fig.~\ref{fig:example_trajectory}, initially $d <d_s$ and also $|\chi-\chi^d(d)| > \pi/2$, i.e. the geometry is in Case - 1. As stated in Remark \ref{reamrk1}, $\vert d \vert$ increases initially as shown in Fig.~\ref{fig:example_d_chi_tilde}. However, from Fig.~\ref{fig:example_chi_tilde}, observe that $|\tilde{\chi}|=|\chi-(\chi^d(d)+\pi/2)|$ monotonically decreases as discussed in proposition \ref{prop_covergence_of_X_line_phase1}. As it reaches near zero, the 1st switching happens as shown as Switch 1 in Fig.~\ref{fig:example}. At Switch 1, following Case - 2 geometry, the vector field has $\Tilde{\chi}=\chi-\chi^d(d)$, which leads to a jump in $\Tilde{\chi}$ profile (see Fig.~\ref{fig:example_chi_tilde}). Then, in line with Proposition \ref{prop_covergence_of_X_Tilde_phase2} for Case - 2 geometry, $|\Tilde{\chi}|$ monotonically decreases. After the transition from Case - 1 to Case - 2, initially $\Dot{|d|}$ is still greater than zero. As a result, $|d|$ increases initially in Case - 2 geometry. However, as $\chi$ crosses $\chi^p$, then $|d|$ starts decreasing (see Fig.~\ref{fig:example_d_chi_tilde}) as per Remark \ref{remark2}. As $\vert d \vert$ crosses $d_s$ in finite time (Proposition \ref{prop_covergence_of_d_for_phase_2}), Switch 2 happens, and the geometry falls in Case - 3, wherein following Proposition \ref{prop_covergence_of_X_Tilde_d_phase3}, $\Tilde{\chi}$ and $\vert d \vert$ monotonically decrease. Besides, as mentioned in Remarks \ref{remark:chi_chattering} and \ref{remark:d_chattering}, no undesirable chattering is observed at the phase transitions. Now, from $\dot{\chi}$ profile in Fig.~\ref{fig:example_chi_d_and_chi_tilde}, $|\Dot{\chi}|_{max}=0.7$ rad/sec, which is within kino-dynamically feasible range. This leads to RHS of \eqref{eq:kappa_max_final} as $\kappa_{max}=|\Dot{\chi}|_{max}/V_a=0.046$~/m. And for the considered reference path, $|\Dot{\chi^p}|_{max}=0.1$~rad/sec. Therefore, LHS of \eqref{eq:kappa_max_final} becomes $0.036$, which satisfies \eqref{eq:kappa_max_final}.

The developed guidance algorithm (a) is now compared with (b) vector field method \cite{nelson2007}, (c) PLOS mehtod \cite{kothari2014uav}, and (d) NLGL ($L_1$) method \cite{park2007},  with parameters fine tuned for better performance as follows. For method (b) $k= 0.02$ and $\beta=\pi/2$, for (c) $K_1=15$ and $K_2=0.1$, and for (d) $L_1=110$~m.  
The initial conditions considered are $\vert d_0 \vert = 200$~m (for (a), (b) and (c)), and $\vert d_0 \vert = 80$~m(for (d)) and $\chi_0 = -\pi/4$ (for all methods). 
% However, for the NLGL algorithm when the initial distance is higher than the $L_1$ value the UAV will not be able to converge, therefore, 
% For the NLGL algorithm the initial conditions are $\vert d_0 \vert = 80$~m and $\chi_0 = -\pi/4$ rad. 
From Figs.~\ref{fig:comparison_trajectories} and ~\ref{fig:comparison_turn_rate}, note that Methods (b) and (c) closely follow the path, but initial turn rate command is very high for method (c), and the convergence time for method (b) is high. On the other hand, method (d) has always higher cross-track error. Unlike them, the proposed switched vector field method allows the UAV to closely follow the path with feasible turn rate and low convergence time. %While the PLOS method initially generates high guidance commands due to error proportionality, as seen in Fig.~\ref{fig:comparison_turn_rate}. on the other hand, the NLGL method produces lower commands but leads to trajectory deviations from the reference path.

Total 200 random trials are next performed with $d_0 \sim \mathcal{U}[100,200]$~m, $\chi_0 \sim \mathcal{U}[-\pi,\pi]$~rad, wind speed $, \vert W \vert \sim \mathcal{U}[2,3]$~m/s, wind direction $\sim \mathcal{U}[-2,-2.5]$~rad and control parameters same as previous simulations. The performance measures are: %were compared with (b), (c) and (d),
% \cite{nelson2007},\cite{kothari2014uav} and \cite{park2007}. 
i) Reaching time ($t_{conv}$), ii) RMS cross track error($d_{rms}$), iii)RMS turn rate, $\Dot{\chi}_{rms}$ and iv) maximum turn rate, max($|\Dot{\chi}|$), as shown in Fig.~\ref{fig:boxplot}. Here, $t_{conv}$ is defined as the duration by which the UAV reaches a threshold distance relative to the reference path, while subsequently following a course angle sufficiently close to path's curvature. While the NLGL method \cite{park2007} exhibits low $\Dot{\chi}_{rms}$, its $d_{rms}$ and $t_{conv}$ are comparatively higher when compared to other methods. For PLOS method \cite{kothari2014uav}, $d_{rms}$ is lower, but it has higher $\Dot{\chi}_{rms}$, $| \Dot{\chi}|_{max}$ and $t_{conv}$. Although the $d_{rms}$ is almost similar for proposed switching vector field-based and the basic Vector field method \cite{nelson2007}, it is evident that the switched vector field-based approach significantly outperforms path-following methods in terms of all performance measures. % both $t_{conv}$, $\Dot{\chi}_{rms}$ and $\vert \Dot{\chi}|_{max}$.

\section{CONCLUSIONS}
\label{sec:conclusions}
This paper has presented a cross-track-error and course angle-dependent novel switching strategy in defining the desired vector field for following a general reference path. Theoretical analysis on its convergence has been presented. The developed guidance scheme has been shown to ensure faster convergence, thus leading to a lesser reaching time, while it also ensures that the guidance command does not go beyond a threshold even when the course angle error is very high. Undesirable chattering at phase transitions is also avoided. Numerical simulation studies have been presented to justify the performance of the developed guidance algorithm. Comparison studies have shown the superiority of the presented method over other existing methods. 
% Future scope of research includes extending the proposed method to following reference circular path and any arbitrary reference path in both 2-D and 3-D environments.
% \begin{figure}[ht]
% \begin{minipage}[b]{0.48\linewidth}
% \centering
% \includegraphics[width=4cm,height=3cm]{figures/comp_PVF_VF_PLOS_L1_trajectory.jpg}
% \caption{UAV trajectories}
% \label{fig:figure1}
% \end{minipage}
% \hspace{0.1cm}
% \begin{minipage}[b]{0.48\linewidth}
% \centering
% \includegraphics[width=4cm,height=3cm]{figures/comp_PVF_VF_PLOS_L1_chi_dot.jpg}
% \caption{urn rate command}
% \label{fig:figure2}
% \end{minipage}
% \end{figure}
% \FloatBarrier

% \section*{ACKNOWLEDGMENT}
% \cite
% \FloatBarrier
%%%%%%%%%%%%%%%%%%%%%%%%%%%%%%%%%%%%%%%%%%%%%%%%%%%%%%%%%%%%%%%%%%%%%%%%%%%%%%%%
\bibliographystyle{IEEEtran}
\bibliography{sample1}

\end{document}